\documentclass[prd,twocolumn,floatfix,amsmath,nofootinbib,amssymb,floatfix]{revtex4}
\usepackage{graphicx,color,dcolumn,booktabs,bm}
\usepackage{longtable,lscape}
\usepackage{pdfpages}
\usepackage{txfonts}
\usepackage{overpic}
\usepackage{amssymb}
\usepackage{makecell}
\usepackage{indentfirst}
\usepackage{feynmf}   %{feynmp}
\usepackage{slashed}  %for Feynman symbols
\usepackage{cases}
\usepackage{color}
\usepackage{multirow}
\usepackage{threeparttable}
\usepackage{epstopdf}
\usepackage{enumerate}
\usepackage{subfigure}
\usepackage{diagbox}
\usepackage{graphicx,color,dcolumn,booktabs,bm}
\usepackage{mathrsfs}
\usepackage{cancel}
\usepackage{float}
\usepackage[colorlinks,
            citecolor=blue,
            anchorcolor=red,
            menucolor=red,
            linkcolor=red,
            filecolor=red,
            runcolor=red,
            urlcolor=blue,
            frenchlinks=red]{hyperref}

\begin{document}
\title{Analysis of the electromagnetic form factors and the radiative decays of the vector heavy-light mesons}
\author{Jie Lu$^{1,2}$}
\author{Guo-Liang Yu$^{1,2}$}
\email{yuguoliang2011@163.com}
\author{Zhi-Gang Wang$^{1}$}
\email{zgwang@aliyun.com}
\author{Bin Wu$^{1}$}

\affiliation{$^1$ Department of Mathematics and Physics, North China
Electric Power University, Baoding 071003, People's Republic of
China\\$^2$ Hebei Key Laboratory of Physics and Energy Technology, North China Electric Power University, Baoding 071000, China}
\date{\today}

\begin{abstract}
In this article, we analyze the electromagnetic form factors of the vector heavy-light mesons to the pseudoscalar heavy-light mesons in the framework of three-point QCD sum rules, where the contributions of vacuum condensate terms $\langle\overline{q}q\rangle$, $\langle\overline{q}g_{s}\sigma Gq\rangle$, $\langle g_{s}^{2}G^{2}\rangle$, $\langle f^{3}G^{3}\rangle$ and $\langle\overline{q}q\rangle\langle g_{s}^{2}G^{2}\rangle$ are considered. With these results, we also obtain the radiative decay widths of the vector heavy-light mesons and then compare our results with those of other collaboration's. The final results about the radiative decay widths are $\Gamma(D^{*0}\to D^{0}\gamma)=1.74^{+0.40}_{-0.37}$ keV, $\Gamma(D^{*+}\to D^{+}\gamma)=0.17^{+0.08}_{-0.07}$ keV, $\Gamma(D_{s}^{*}\to D_{s}\gamma)=0.029^{+0.009}_{-0.008}$ keV, $\Gamma(B^{*0}\to B^{0}\gamma)=0.018^{+0.006}_{-0.005}$ keV, $\Gamma(B^{*+}\to B^{+}\gamma)=0.015^{+0.007}_{-0.007}$ keV and $\Gamma(B^{*}_{s}\to B_{s}\gamma)=0.016^{+0.003}_{-0.005}$ keV.
\end{abstract}


\maketitle

\section{Introduction}\label{sec1}

As one of the most interesting research areas in particle physics, the experimental and theoretical investigations about the heavy flavor hadrons have been developing rapidly. This is due to their key roles in both understanding of QCD long-distance dynamics, and the determination of the fundamental parameters of the standard model (SM). Especially, the decay behavior of heavy flavor hadrons is an important and interesting research topic for testing the SM and finding new physics beyond the SM. These decay processes can be classified into leptonic, semileptonic and nonleptonic decays which involve both electroweak and strong interactions. However, theoretical investigations about the heavy hadrons is very difficult because the QCD is non-perturbative in low energy regions. Therefore, many phenomenological approaches\cite{Beneke:1999br,Muta:2000ti,Lu:2000em,Keum:2000ph,Keum:2000wi,Lu:2000hj,Bauer:2000yr,Bauer:2001yt,FermilabLattice:2011njy,Yang:2005bv,Aliev:2007rq,Wang:2007ys,Ghahramany:2009zz,Wang:2008da,Colangelo:2010bg,Sun:2010nv,Han:2013zg,Colangelo:1994jc,Cho:1992nt,Ball:1997rj,Casalbuoni:1996pg,Deandrea:1998uz,Orsland:1998de,Goity:2000dk,Li:2008tk,Chen:2007na,Choi:2007se,Jaus:1996np,Wang:2019mhm,Li:2022vby,Lu:2022kos,Cui:2023jiw,Ivanov:2022nnq,Tran:2023hrn} have been employed to analyze the decay processes of the heavy flavor hadrons.

The QCD sum rules (QCDSR) is one of the most powerful non-perturbative approach, and has been widely used to analyze the mass spectra and the decay behavior of hadrons \cite{Lucha:2011zp,Narison:2012xy,Gelhausen:2013wia,Zhu:1996qy,Zhu:2000py,Wang:2010it,Chen:2016phw,Azizi:2020tgh,Duraes:2007te,Liu:2007fg,Zhang:2008pm,Xin:2021wcr,Li:2015xka,Yu:2019sqp,Wang:2023sii,Wang:2023kir}. In recent years, some tasks were carried out by three-point QCDSR, such as the analysis of electroweak and electromagnetic form factors\cite{Wang:2007ys,Aliev:1994nq,Khosravi:2013lea,Peng:2019apl,Shi:2019hbf,Zhao:2020mod,Xing:2021enr,Zhang:2023nxl}, and the strong coupling constants \cite{Azizi:2010jj,Sundu:2011vz,Cui:2012wk,Bracco:2011pg,Rodrigues:2017qsm,Lu:2023gmd,Lu:2023lvu,Azizi:2014bua,Azizi:2015tya,Yu:2018hnv,Lu:2023pcg}. These parameters are very important to analyze the decay process of hadrons. In our previous work, the $B_{c}^{*}\to B_{c}$ electromagnetic form factor were studied by three-point QCDSR, and the corresponding radiative decay $B_{c}^{*}\to B_{c}\gamma$ was calculated\cite{Wang:2013cha}. In Ref\cite{Yu:2015xwa}, we analyzed the momentum dependent strong coupling constant $G_{D_{s}^{*}D_{s}\phi}(Q^{2})$. According vector meson dominant(VMD), the electromagnetic coupling constant $G_{D_{s}^{*}D_{s}\gamma}$ was also obtained by setting $Q^{2}=0$. And the radiative decay width for $D_{s}^{*}\to D_{s}\gamma$ was determined to be $0.59\pm0.15$ keV. As a continuation of these works, we systematically analyze the electromagnetic form factors of vector heavy-light mesons to pseudoscalar heavy-light mesons. Based on these results, the decay widths about vector heavy-light mesons to pseudoscalar heavy-light mesons plus a photon are obtained. Although some radiative decays, such as $D^{*+}\to D^{+}\gamma$, $D^{*0}\to D^{0}\gamma$ and $D_{s}^{*}\to D_{s}\gamma$ were already analyzed by other research groups\cite{Casalbuoni:1996pg,Deandrea:1998uz,Orsland:1998de,Goity:2000dk,Aliev:1994nq}. However, these results were obtained by different methods and not consistent well with each other, which needs further confirmation by other theoretical methods.

The experimental data for the radiative decay of the vector meson $D^{*+}$ in Particle Data Group(PDG)\cite{ParticleDataGroup:2022pth} is $\Gamma_{Total}(D^{*+})=83.4\pm1.8$ keV, and $\frac{\Gamma(D^{*+}\to D^{+}\gamma)}{\Gamma_{Total}(D^{*+})}\approx 1.6\pm0.4\%$. The decay widths for $D^{*0}$ and $D_{s}^{*}$ are given as $\Gamma_{Total}(D^{*0})<2.1$ MeV, $\frac{\Gamma(D^{*0}\to D^{0}\gamma)}{\Gamma_{Total}(D^{*0})}\approx 35.3\pm0.9\%$, and $\Gamma_{Total}(D_{s}^{*})<1.9$ MeV, $\frac{\Gamma(D_{s}^{*}\to D_{s}\gamma)}{\Gamma_{Total}(D^{*}_{s})}\approx(93.5\pm0.7)\%$, respectively. However, more exact values about these decays have not been determined yet. For the mesons $B^{*\pm(0)}$ and $B_{s}^{*}$, the radiative decays as their main decay channels have not been determined in experiment. In general, more exact experimental results and theoretical predictions need to be provided, which can help us to understand the nature of the mesons and test the theoretical models.

The layout of this paper is as follows, After introduction in Sec. \ref{sec1}, the radiative decays of the vector heavy-light mesons are analyzed in the framework of SM in Sec. \ref{sec2}, and the electromagnetic form factor is introduced. In Sec. \ref{sec3}, we systematically analyze the electromagnetic form factors of vector heavy-light meson to pseudoscalar heavy-light meson by the three-point QCDSR, where the contributions of perturbative part and vacuum condensate including $\langle\overline{q}q\rangle$, $\langle\overline{q}g_{s}\sigma Gq\rangle$, $\langle g_{s}^2G^{2} \rangle$, $\langle f^{3}G^{3}\rangle$ and $\langle\overline{q}q\rangle\langle g_{s}^{2}G^{2}\rangle$ are considered in QCD side. Sec. \ref{sec4} is employed to present the numerical results and discussions. Sec. \ref{sec5} is reserved as conclusions. Some important figures are shown in Appendix.

\section{The radiative decay of vector heavy-light mesons}\label{sec2}

To analyze the radiative decay of vector heavy-light mesons($\mathbb{V}$) to pseudoscalar heavy-light mesons($\mathbb{P}$) ( $\mathbb{V}\to\mathbb{P}\gamma$), we firstly write the following electromagnetic lagrangian $\mathscr{L}_{EM}$,
\begin{eqnarray}\label{eq:1}
\mathscr{L}_{EM}=-e_{Q}\bar{Q}\gamma_{\mu}QA^{\mu}-e_{q}\bar{q}\gamma_{\mu}qA^{\mu}
\end{eqnarray}
where $A^{\mu}$ is the electromagnetic field, $q$ and $Q$ denote the light (u, d or s) and heavy (c, b) quarks. $e_{q}$ and $e_{Q}$ are electric charges which are taken as $-\frac{1}{3}e$ for d, s and b quarks, and $\frac{2}{3}e$ for u and c quarks. The assignments of $\mathbb{P}$, $\mathbb{V}$, $Q$ and $q$ for different decay modes are shown in Table~\ref{AM}.
\begin{table}[htbp]
\caption{The assignments of $\mathbb{P}$, $\mathbb{V}$, $Q$ and $q$ for different decay modes.}
\label{AM}
\begin{tabular}{p{1.6cm}<{\centering} p{1.2cm}<{\centering} p{1.2cm}<{\centering} p{1.2cm}<{\centering} p{1.2cm}<{\centering} }
\hline
\hline
Mode&$\mathbb P$&$\mathbb V$&$Q$&$q$ \\ \hline
$D^{*+}\to D^{+}\gamma$&$D^{+}$&$D^{*+}$&c&d \\
$D^{*0}\to D^{0}\gamma$&$D^{0}$&$D^{*0}$&c&u \\
$D_{s}^{*+}\to D^{+}_{s}\gamma$&$D_{s}^{+}$&$D_{s}^{*+}$&c&s \\
$B^{*-}\to B^{-}\gamma$&$B^{-}$&$B^{*-}$&b&u \\
$\bar{B}^{*0}\to \bar{B}^{0}\gamma$&$\bar{B}^{0}$&$\bar{B}^{*0}$&b&d \\
$B_{s}^{*0}\to B_{s}^{0}\gamma$&$B_{s}^{0}$&$B_{s}^{*0}$&b&s \\ \hline\hline
\end{tabular}
\end{table}

\begin{figure}[htbp]
\centering
\subfigure[]{\includegraphics[width=4cm]{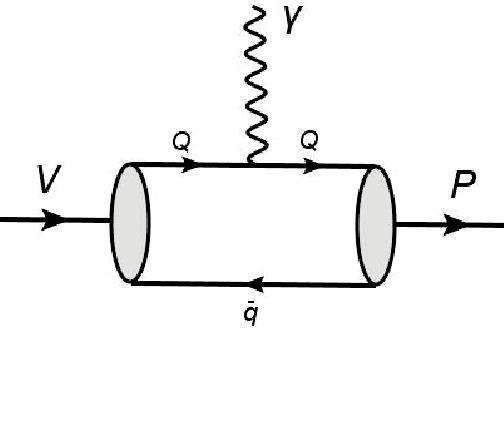}}
\subfigure[]{\includegraphics[width=4cm]{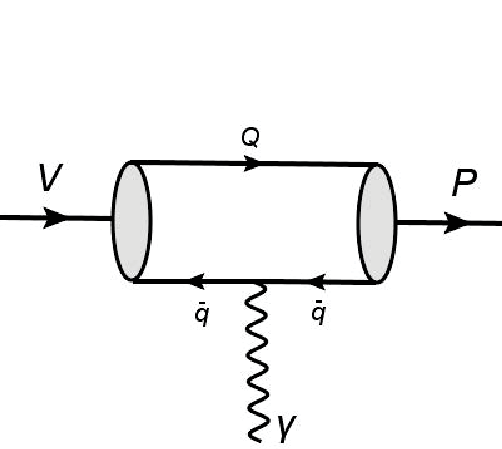}}
\caption{The feynman diagrams of radiative decay for $\mathbb{V}\to\mathbb{P}\gamma$.}
\label{HLFD}
\end{figure}

From this lagrangian(Eq. (\ref{eq:1})), the decay amplitude can be expressed as the following form,
\begin{eqnarray}\label{eq:2}
\notag
&&\left\langle {\mathbb P(p')\gamma (q)} \right|{\mathscr{L}_{EM}}\left| {\mathbb V(p)} \right\rangle \\
\notag
&& = \left\langle {\mathbb P(p')\gamma (q)} \right|( - {e_Q}\bar Q{\gamma _\mu }Q - {e_q}\bar q{\gamma _\mu }q){A^\mu }\left| {\mathbb V(p)} \right\rangle \\
&& = -\left\langle {\gamma (q)} \right|{A^\mu }\left| 0 \right\rangle \left\langle {\mathbb P(p')} \right|( {e_Q}\bar Q{\gamma _\mu }Q + {e_q}\bar q{\gamma _\mu }q)\left| {\mathbb V(p)} \right\rangle
\end{eqnarray}
with
\begin{eqnarray}\label{eq:3}
\notag
&&\left\langle {\mathbb{P}(p')} \right|{e_Q}\bar Q{\gamma _\mu }Q + {e_q}\bar q{\gamma _\mu }q\left| {\mathbb{V}(p)} \right\rangle  \\
&& = {\varepsilon _{\mu \sigma \alpha \beta }}{\zeta ^\sigma }{p^\alpha }p{'^\beta }\frac{{V({q^2})}}{{{m_\mathbb V} + {m_\mathbb P}}}
\end{eqnarray}
where $V(q^{2})$ is electromagnetic form factor, $p^{\alpha}$ and $p'^{\beta}$ are four momentum of the vector and pseudoscalar heavy-light mesons, and $\zeta^{\sigma}$ is the polarization vector of vector heavy-light mesons. According to Eqs. (\ref{eq:2}) and (\ref{eq:3}), the decay amplitude finally can be written as,
\begin{eqnarray}\label{eq:4}
\notag
&&\left\langle {\mathbb P(p')\gamma (q)} \right|{\mathscr{L}_{EM}}\left| {\mathbb V(p)} \right\rangle \\
&& = {\varepsilon _{\mu \sigma \alpha \beta }}{\zeta ^\sigma }{p^\alpha }p{'^\beta }\frac{{V(0)}}{{{m_\mathbb V} + {m_\mathbb P}}}( - i){\eta ^\mu }
\end{eqnarray}
where $\eta_{\mu}$ is the polarization vector of the photon. This decay process can be explicitly illustrated by feynman diagrams in Fig. \ref{HLFD}.

\section{QCDSR for the electromagnetic form factors}\label{sec3}

To study the electromagnetic form factor of the vector heavy-light meson to the pseudoscalar heavy-light meson, we firstly write down the three-point correlation function,
\begin{eqnarray}\label{eq:5}
\notag
{\Pi _{\mu \nu }}(p,p') &&= {i^2}\int {{d^4}x{d^4}y{e^{ip'x}}{e^{i(p - p')y}}} \\
&& \times \left\langle 0 \right|T\{ {J_{\mathbb{P}}}(x){j_\mu }(y)J_\nu ^ + (0)\} \left| 0 \right\rangle
\end{eqnarray}
where $T$ denotes the time ordered product, $J_{\mathbb{P}}$ and $J^{+}_{\nu}$ are the interpolating currents of the pseudoscalar and vector heavy-light mesons, and $j_{\mu}$ is the electromagnetic current. These currents are taken as the following forms,
\begin{eqnarray}\label{eq:6}
\notag
{J_\mathbb P}(x) &&= \bar q(x)i{\gamma _5}Q(x)\\
\notag
{j_\mu }(y) &&= {{e_Q}\bar Q(y){\gamma _\mu }} Q(y) + {{e_q}\bar q(y){\gamma _\mu }} q(y) \\
{J_\nu }(0) &&= \bar q(0){\gamma _\nu }Q(0)
\end{eqnarray}

In the framework of QCDSR, the correlation function will be calculated at two sides which are called the phenomenological side and the QCD side. Then, according to the quark hadron duality, the calculations of these two sides will be coordinated and the sum rules for electromagnetic form factor in the hadronic level can be obtained.

\subsection{The Phenomenological side}\label{sec3.1}

In the phenomenological side, a complete sets of hadronic states with the same quantum numbers as the interpolating currents $J_{\mathbb{P}}$ and $J^{+}_{\nu}$ are inserted into the correlation function. By using the dispersion relation, the correlation function can be written as\cite{Colangelo:2000dp},
\begin{eqnarray}
\notag
{\Pi^{\mathrm{phy}} _{\mu \nu }}(p,p') &&= \frac{{\left\langle 0 \right|{J_\mathbb P}(0)\left| {\mathbb P(p')} \right\rangle \left\langle {\mathbb V(p)} \right|J_\nu ^ + (0)\left| 0 \right\rangle }}{{(m_\mathbb V^2 - {p^2})(m_\mathbb P^2 - p{'^2})}}\\
&& \times \left\langle {\mathbb P(p')} \right|{j_\mu }(y)\left| {\mathbb V(p)} \right\rangle +h.c.
\end{eqnarray}
where $h.c.$ represents the contributions coming from higher resonances and continuum states. The hadronic matrix elements are defined by the following parameterized equations,
\begin{eqnarray}
\notag
\left\langle 0 \right|{J_\mathbb P}(0)\left| {\mathbb P(p')} \right\rangle  &&= \frac{{{f_\mathbb P}m_\mathbb P^2}}{{{m_q} + {m_Q}}}\\
\notag
\left\langle {\mathbb V(p)} \right|J_\nu ^ + (0)\left| 0 \right\rangle  &&= {f_\mathbb V}{m_\mathbb V}\zeta _\nu ^*\\
\left\langle {\mathbb P(p')} \right|{j_\mu }(y)\left| {\mathbb V(p)} \right\rangle  &&= {\varepsilon _{\mu \sigma \alpha \beta }}{\zeta ^\sigma }{p^\alpha }p{'^\beta }\frac{{V({q^2})}}{{{m_\mathbb V} + {m_\mathbb P}}}
\end{eqnarray}
with $q=p-p'$. In these above equations, $f_{\mathbb{P}}$ and $f_{\mathbb{V}}$ are the decay constants of the pseudoscalar and vector heavy-light mesons, $\varepsilon_{\mu\sigma\alpha\beta}$ is the Levi-Civita tensor, and $V(q^{2})$ is the electromagnetic form factor.  $\zeta_{\nu}$ is the polarization vector of vector heavy-light meson with the following properties,
\begin{eqnarray}
\zeta _\nu ^*{\zeta _\sigma } =  - {g_{\nu \sigma }} + \frac{{{p_\nu }{p_\sigma }}}{{{p^2}}}
\end{eqnarray}
With these above equations, the correlation function in phenomenological side can be expressed as the following form,
\begin{eqnarray}
\notag
{\Pi^{\mathrm{phy}} _{\mu \nu }}(p,p') &&=  - \frac{{{f_\mathbb P}m_\mathbb P^2{f_\mathbb V}{m_\mathbb V}}}{{({m_q} + {m_Q})({m_\mathbb V} + {m_\mathbb P})}}\\
\notag
&& \times \frac{{V({q^2}){\varepsilon _{\mu \nu \alpha \beta }}{p^\alpha }p{'^\beta }}}{{(m_\mathbb V^2 - {p^2})(m_\mathbb P^2 - p{'^2})}} + h.c.\\
&&=\Pi^{\mathrm{phy}}(p,p')\varepsilon _{\mu \nu \alpha \beta}{p^\alpha }p{'^\beta }+h.c.
\end{eqnarray}
where $\Pi^{\mathrm{phy}}(p,p')$ is called scalar invariant amplitude, $\varepsilon_{\mu\nu\alpha\beta}p^{\alpha}p'^{\beta}$ is the corresponding tensor structure.

\subsection{The QCD side}\label{sec3.2}
In this part, we firstly contract the quark filed with Wick's theorem, and then do the operator product expansion(OPE). After contraction of the quark filed, the correlation function in QCD side can be expressed as,
\begin{eqnarray}\label{eq:11}
\notag
{\Pi^{\mathrm{QCD}} _{\mu \nu }}(p,p') &&=  - i\int {{d^4}x{d^4}y{e^{ip'x}}{e^{i(p - p')y}}} \\
\notag
&& \times \{ {e_Q}Tr[S_Q^{nk}(y){\gamma _\nu }S_q^{km}( - x){\gamma _5}S_Q^{mn}(x - y){\gamma _\mu }]\\
\notag
&& + {e_q}Tr[S_Q^{nk}(x){\gamma _\nu }S_q^{km}( - y){\gamma _\mu }S_q^{mn}(y - x){\gamma _5}]\} \\
&& =e_{Q} {\Pi _{1\mu \nu }}(p,p') + e_{q} {\Pi _{2\mu \nu }}(p,p')
\end{eqnarray}
where $S_{q}^{mn}(x)$ and $S_{Q}^{mn}(x)$ are the full propagators of light and heavy quarks, and they have the following forms\cite{Pascual:1984zb,Reinders:1984sr},
\begin{eqnarray}
\notag
{S_q^{mn}}(x) &&= \frac{{i{\delta ^{mn}}x\!\!\!/}}{{2{\pi ^2}{x^4}}} - \frac{{{\delta ^{mn}}{m_q}}}{{4{\pi ^2}{x^4}}} - \frac{{{\delta ^{mn}}\left\langle {\bar qq} \right\rangle }}{{12}} + \frac{{i{\delta ^{mn}}x\!\!\!/{m_q}\left\langle {\bar qq} \right\rangle }}{{48}}\\
\notag
&& - \frac{{{\delta ^{mn}}{x^2}\left\langle {\bar q{g_s}\sigma Gq} \right\rangle }}{{192}} + \frac{{i{\delta ^{mn}}{x^2}x\!\!\!/{m_q}\left\langle {\bar q{g_s}\sigma Gq} \right\rangle }}{{1152}}\\
\notag
&& - \frac{{i{g_s}G_{\alpha \beta }^at_{mn}^a(x\!\!\!/{\sigma ^{\alpha \beta }} + {\sigma ^{\alpha \beta }}x\!\!\!/)}}{{32{\pi ^2}{x^2}}} - \frac{{i{\delta ^{mn}}{x^2}x\!\!\!/g_s^2{{\left\langle {\bar qq} \right\rangle }^2}}}{{7776}}\\
\notag
&& - \frac{{{\delta ^{mn}}{x^4}\left\langle {\bar qq} \right\rangle \left\langle {g_s^2GG} \right\rangle }}{{27648}} - \frac{{\left\langle {{{\bar q}^n}{\sigma ^{\mu \nu }}{q^m}} \right\rangle {\sigma _{\mu \nu }}}}{8}\\
\notag
&& - \frac{{\left\langle {{{\bar q}^n}{\gamma ^\mu }{q^m}} \right\rangle {\gamma _\mu }}}{4} + ...\\
\notag
{S_Q^{mn}}(x) &&= \frac{i}{{{{(2\pi )}^4}}}\int {{d^4}k} {e^{ - ik \cdot x}}\{ \frac{{{\delta ^{mn}}}}{{k\!\!\!/ - {m_Q}}}\\
\notag
&& - \frac{{{g_s}G_{\alpha \beta }^at_{mn}^a}}{4}\frac{{{\sigma ^{\alpha \beta }}(k\!\!\!/ + {m_Q}) + (k\!\!\!/ + {m_b}){\sigma ^{\alpha \beta }}}}{{{{({k^2} - m_Q^2)}^2}}}\\
\notag
&& + \frac{{{g_s}{D_\alpha }G_{\beta \lambda }^at_{mn}^a({f^{\lambda \beta \alpha }} + {f^{\lambda \alpha \beta }})}}{{3{{({k^2} - m_Q^2)}^4}}}\\
\notag
&& - \frac{{g_s^2{{({t^a}{t^b})}_{mn}}G_{\alpha \beta }^aG_{\mu \nu }^b({f^{\alpha \beta \mu \nu }} + {f^{\alpha \mu \beta \nu }} + {f^{\alpha \mu \nu \beta }})}}{{4{{({k^2} - m_Q^2)}^5}}} \\
&&+ ...\}
\end{eqnarray}
Here, $\langle g_{s}^{2}GG\rangle=\langle g_{s}^{2}G^{a}_{\alpha\beta}G^{a\alpha\beta}\rangle$, $D_{\alpha}=\partial_{\alpha}-ig_{s}G_{\alpha}^{a}t^{a}$, $t^{a}=\frac{\lambda^{a}}{2}$. $\lambda^{a}$($a$=1,...,8) are the Gell-Mann matrices, $m$ and $n$ are color indices, $q$ denotes the light quarks(u, d, and s), $Q$ represents the heavy quarks(c and b), $\sigma_{\alpha\beta}=\frac{i}{2}[\gamma_{\alpha},\gamma_{\beta}]$, and $f^{\lambda\alpha\beta}$, $f_{\alpha\beta\mu\nu}$ have the following forms,
\begin{eqnarray}
{f^{\lambda \alpha \beta }} = (k\!\!\!/ + {m_Q}){\gamma ^\lambda }(k\!\!\!/ + {m_Q}){\gamma ^\alpha }(k\!\!\!/ + {m_Q}){\gamma ^\beta }(k\!\!\!/ + {m_Q})
\end{eqnarray}
\begin{eqnarray}\label{eq:10}
\notag
{f^{\alpha \beta \mu \nu }} = && (k\!\!\!/ + {m_Q}){\gamma ^\alpha }(k\!\!\!/ + {m_Q}){\gamma ^\beta }(k\!\!\!/ + {m_Q})\\
&&{\gamma ^\mu }(k\!\!\!/ + {m_Q}){\gamma ^\nu }(k\!\!\!/ + {m_Q})
\end{eqnarray}

From these above equations, the correlation function in QCD side(Eq. (\ref{eq:11})) can also be transformed into the following form,
\begin{eqnarray}
\Pi^{\mathrm{QCD}}_{\mu\nu}(p,p')=\Pi^{\mathrm{QCD}}(p^{2},p'^{2},q^2)\epsilon_{\mu\nu\alpha\beta}p^{\alpha}p'^{\beta}
\end{eqnarray}
where $\Pi^{\mathrm{QCD}}$ is the scalar invariant amplitude in QCD side, and it can be divided into several parts according to different condensate terms,
\begin{eqnarray}
\notag
\Pi^{\mathrm{QCD}}&&=\Pi^{\mathrm{pert}}+\Pi^{\langle\overline{q}q\rangle}+\Pi^{\langle g_{s}^{2}G^{2}\rangle}+\Pi^{\langle\overline{q}g_{s}\sigma Gq\rangle}\\
&&+\Pi^{\langle f^{3}G^{3}\rangle}+\Pi^{\langle\overline{q}q\rangle\langle g_{s}^{2}G^{2}\rangle}+...
\end{eqnarray}
Here, $\Pi^{\mathrm{pert}}$ represents the perturbative part, and $\Pi^{\langle\overline{q}q\rangle}$, $\Pi^{\langle g_{s}^{2}G^{2}\rangle}$, $\Pi^{\langle\overline{q}g_{s}\sigma Gq\rangle}$, $\Pi^{\langle f^3G^3 \rangle}$ and $\Pi^{\langle\overline{q}q\rangle\langle g_{s}^{2}G^{2}\rangle}$ are the vacuum condensate terms with dimension 3, 4, 5, 6 and 7, respectively. The perturbative part and the gluon condensate terms $\Pi^{\langle g_{s}^{2}G^{2}\rangle}$ and $\Pi^{\langle f^3G^3 \rangle}$ can be written as the following form according to the dispersion relation,
\begin{eqnarray}
\notag
\Pi (p,p') =  - \int\limits_{s_1}^\infty  {\int\limits_{u_1}^\infty  {\frac{{{\rho }(s,u,{q^2})}}{{(s - {p^2})(u - p{'^2})}}dsdu}}
\end{eqnarray}
where $\rho(s,u,q^2)$ is the QCD spectral density and it can be represented as,
\begin{eqnarray}
\notag
\rho (s,u,{q^2}) &&= {\rho ^{\mathrm{pert}}}(s,u,{q^2}) + {\rho ^{\left\langle {g_s^2{G^2}} \right\rangle }}(s,u,{q^2}) \\
&&+ {\rho ^{\left\langle {{f^3}{G^3}} \right\rangle }}(s,u,{q^2})
\end{eqnarray}
with $s=p^{2}$, $u=p'^{2}$ and $q=p-p'$.

\begin{figure}[htbp]
\centering
\subfigure[]{\includegraphics[width=4cm]{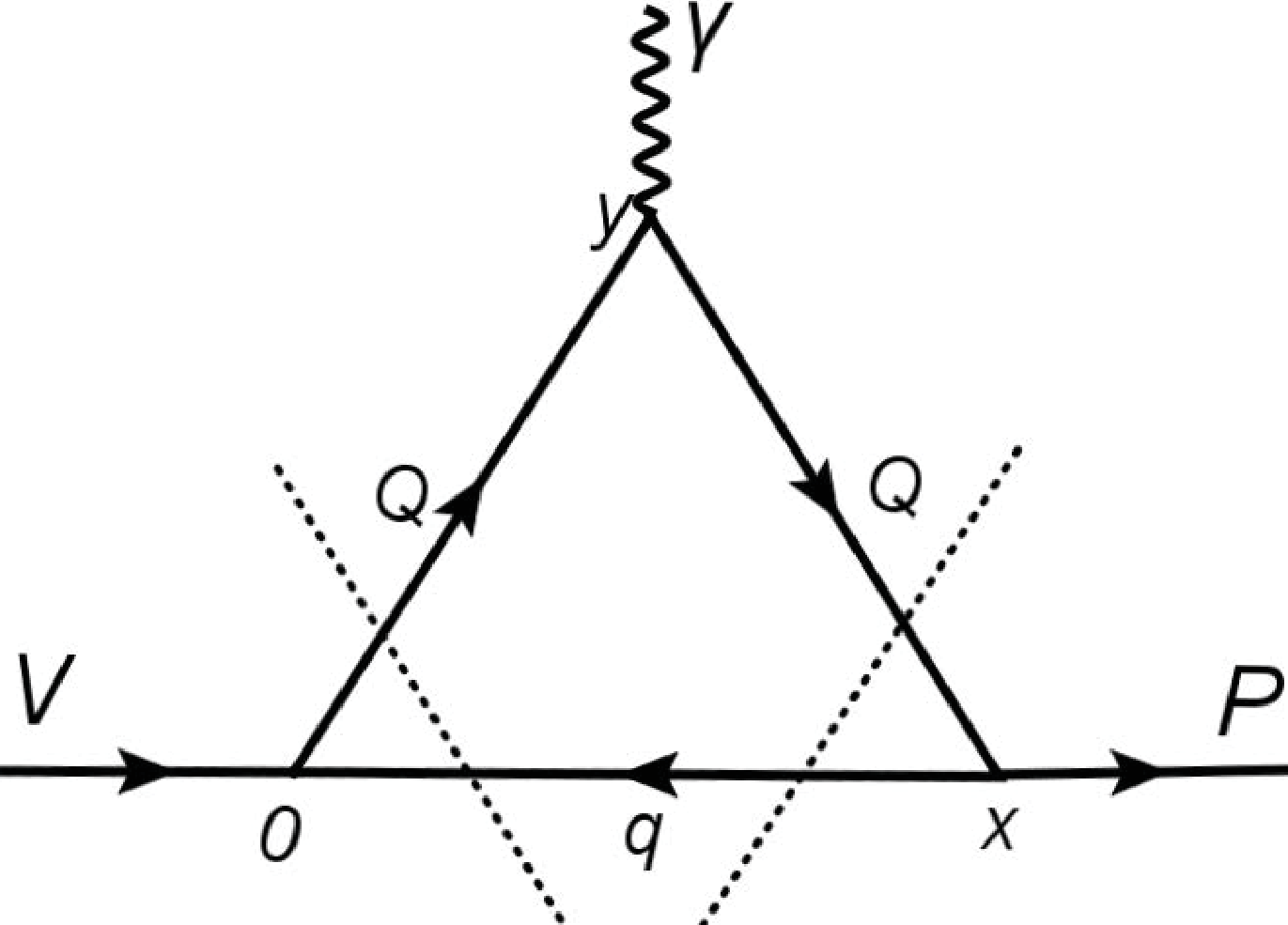}}
\subfigure[]{\includegraphics[width=4cm]{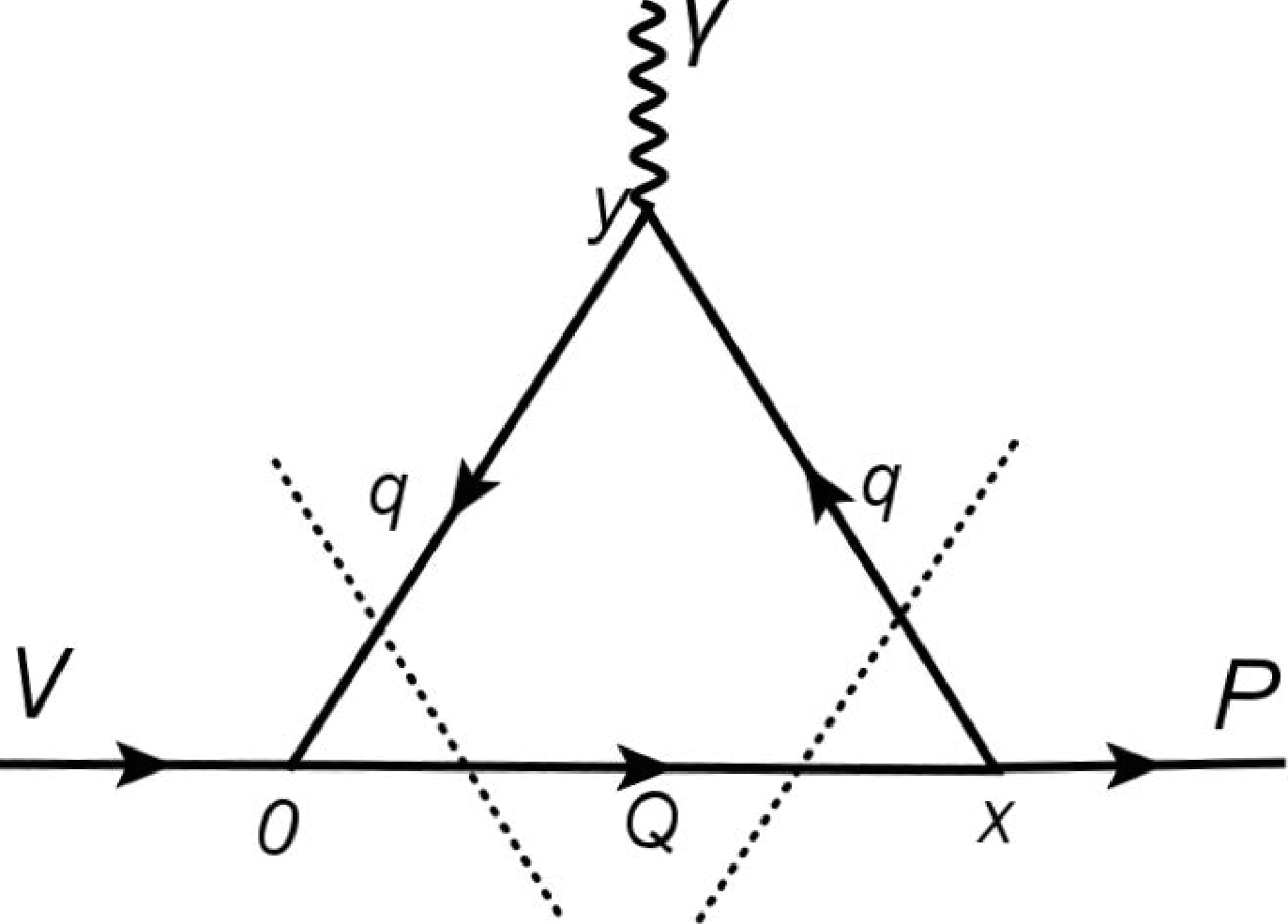}}
\caption{The feynman diagrams for $\Pi^{\mathrm{pert}}_{1}$(a) and $\Pi^{\mathrm{pert}}_{2}$(b) in Eq. (\ref{eq:18}). The dashed lines denote the cutkosky cuts.}
\label{free}
\end{figure}

For the perturbative part, we firstly substitute the free propagator of heavy and light quarks in the momentum space in Eq. (\ref{eq:11}). Then, the correlation function can be expressed as the following form after performing the integrations in the coordinate and momentum space,
\begin{eqnarray}\label{eq:18}
\notag
&&\Pi _{\mu \nu }^{\mathrm{pert}}(p,p') =  - \frac{3}{{{{(2\pi )}^4}}}\int {{d^4}k} \\
\notag
&&\times \{ {e_Q}\frac{{Tr[(\slashed k + \slashed q + {m_Q}){\gamma _\nu }(\slashed k - \slashed p' + {m_q}){\gamma _5}(\slashed k + {m_Q}){\gamma _\mu }]}}{{[{{(k + q)}^2} - m_Q^2][{{(k - p')}^2} - m_q^2]({k^2} - m_Q^2)}}\\
\notag
&&+ {e_q}\frac{{Tr[(\slashed k + \slashed p' + {m_Q}){\gamma _\nu }(\slashed k - \slashed q + {m_q}){\gamma _\mu }(\slashed k + {m_q}){\gamma _5}]}}{{[{{(k + p')}^2} - m_Q^2][{{(k - q)}^2} - m_q^2]({k^2} - m_q^2)}}\} \\
&&= {e_Q}\Pi _{1\mu \nu }^{pert}(p,p') + {e_q}\Pi _{2\mu \nu }^{pert}(p,p')
\end{eqnarray}
Then, the QCD spectral density for the perturbative contribution can be obtained by putting all the quark lines on mass-shell using the Cutkoskys's rules(see Fig. \ref{free}),
\begin{eqnarray}\label{eq:19}
\notag
&&\rho _{1\mu \nu }^{\mathrm{pert}}(s,u,{q^2}) = \frac{3}{{{{(2\pi )}^3}}}\int {{d^4}k\delta } [{(k + q)^2} - m_Q^2]\\
\notag
&& \times \delta [{(k - p')^2} - m_q^2]\delta ({k^2} - m_Q^2)Tr[(\slashed k + \slashed q + {m_Q}){\gamma _\nu }\\
\notag
&& \times (\slashed k - \slashed p' + {m_q}){\gamma _5}(\slashed k + {m_Q}){\gamma _\mu }]\\
\notag
&&\rho _{2\mu \nu }^{\mathrm{pert}}(s,u,{q^2}) =  - \frac{3}{{{{(2\pi )}^3}}}\int {{d^4}k\delta [{{(k - p')}^2} - m_Q^2]} \\
\notag
&& \times \delta [{(k + q)^2} - m_q^2]\delta ({k^2} - m_q^2)Tr[(\slashed k - \slashed p' - {m_Q}){\gamma _\nu }\\
&& \times (\slashed k + \slashed q - {m_q}){\gamma _\mu }(\slashed k - {m_q}){\gamma _5}]
\end{eqnarray}
After performing integrations of these Dirac delta functions in Eq. (\ref{eq:19}), the spectral density can be expressed as,
\begin{eqnarray}
\notag
&&\rho _{1\mu \nu }^{\mathrm{pert}}(s,u,{q^2}) = \frac{3}{{{{(2\pi )}^3}}}\frac{\pi }{{2\sqrt {\lambda (s,u,{q^2})} }}Tr\{ [({C_p} + 1)\slashed{p} \\
\notag
&& + ({C_{p'}} - 1)\slashed{p}'+ {m_Q}){\gamma _\nu }[{C_p}\slashed{p} + ({C_{p'}} - 1)\slashed{p}' + {m_q}]{\gamma _5} \\
\notag
&& \times({C_p}\slashed{p} + {C_{p'}}\slashed{p}' + {m_Q}){\gamma _\mu }\} \\
\notag
&&\rho _{2\mu \nu }^{\mathrm{pert}}(s,u,{q^2}) = -\frac{3}{{{{(2\pi )}^3}}}\frac{\pi }{{2\sqrt {\lambda (s,u,{q^2})} }}Tr\{ [C{'_p}\slashed{p}\\
\notag
&& + (C{'_{p'}} - 1)\slashed{p}'- {m_Q}){\gamma _\nu }[(C{'_p} + 1)\slashed{p} + (C{'_{p'}} - 1)\slashed{p}' \\
&&- {m_q}]{\gamma _\mu }(C{'_p}\slashed{p} + C{'_{p'}}\slashed{p}' - {m_Q}){\gamma _5}\}
\end{eqnarray}
where,
\begin{eqnarray}
\notag
&&{C_p} = \frac{{(u + m_Q^2 - m_q^2)(s + u - {q^2}) - 2u(u - {q^2} + m_Q^2 - m_q^2)}}{{\lambda (s,u,{q^2})}}\\
\notag
&&{C_{p'}} = \frac{{(u - {q^2} + m_Q^2 - m_q^2)(s + u - {q^2}) - 2s(u + m_Q^2 - m_q^2)}}{{\lambda (s,u,{q^2})}}\\
\notag
&&C{'_p} = \frac{{(u + m_q^2 - m_Q^2)(s + u - {q^2}) - 2u(u - {q^2} + m_q^2 - m_Q^2)}}{{\lambda (s,u,{q^2})}}\\
\notag
&&C{'_{p'}} = \frac{{(u - {q^2} + m_q^2 - m_Q^2)(s + u - {q^2}) - 2s(u + m_q^2 - m_Q^2)}}{{\lambda (s,u,{q^2})}}\\
&&\lambda (s,u,{q^2}) = {(s + u - {q^2})^2} - 4su
\end{eqnarray}

\begin{figure*}[htbp]
\centering
\includegraphics[width=16cm]{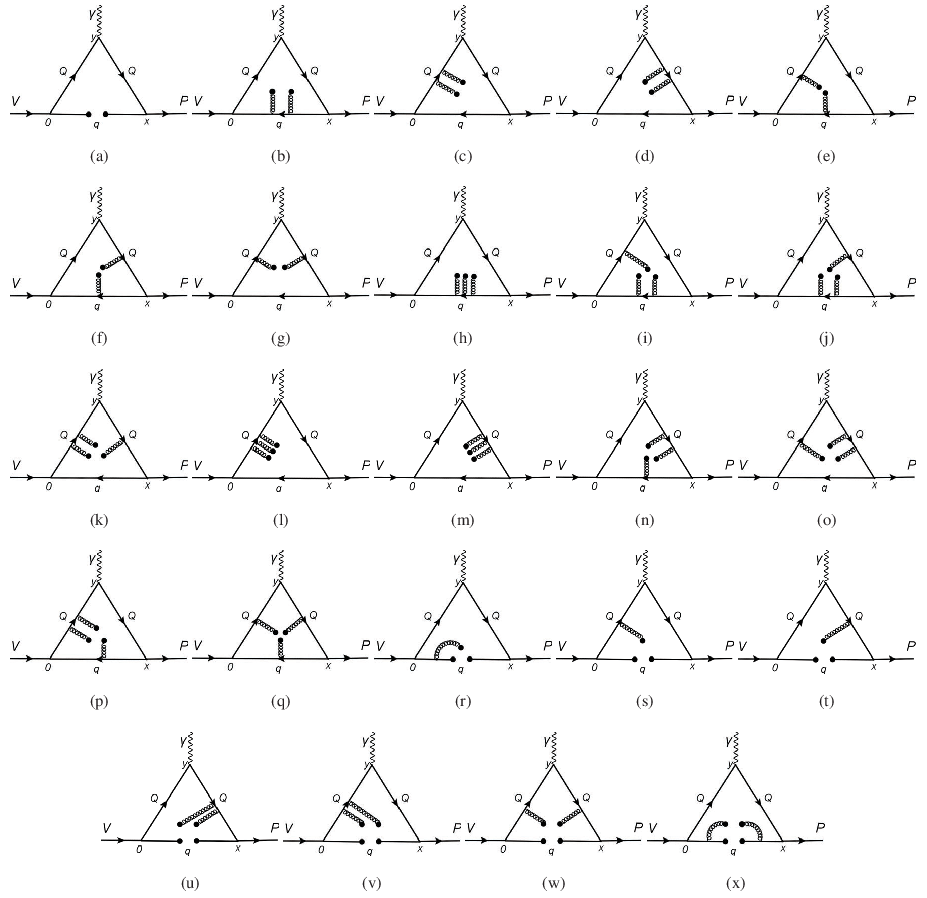}
\caption{Feynman diagrams of the vacuum condensate terms $\langle\overline{q}q\rangle$(a), $\langle g_{s}^{2}G^{2} \rangle$(b-g), $\langle f^{3}G^{3} \rangle$(h-q), $\langle\overline{q}g_{s}\sigma Gq \rangle$(r-t) and $\langle\overline{q}q\rangle\langle g_{s}G^{2} \rangle$(u-x) for $\Pi_{1}$ in Eq. (\ref{eq:11}).}
\label{fig:FM1}
\end{figure*}

\begin{figure*}[htbp]
\centering
\includegraphics[width=16cm]{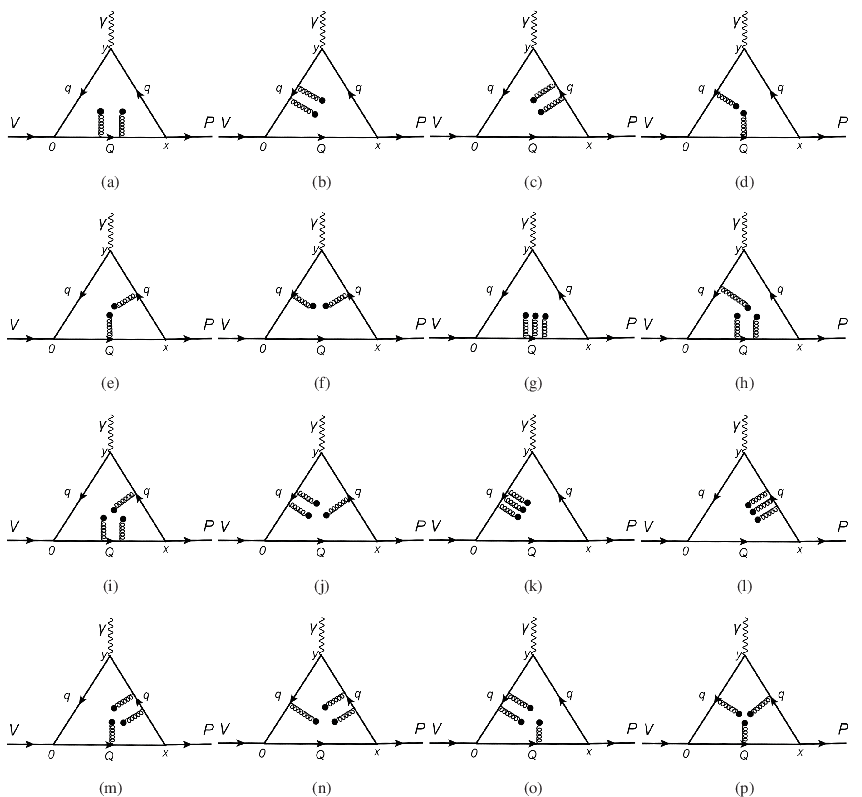}
\caption{Feynman diagrams of the vacuum condensate terms $\langle g_{s}^{2}G^{2} \rangle$(a-f) and $\langle f^{3}G^{3} \rangle$(g-p) for $\Pi_{2}$ in Eq. (\ref{eq:11}).}
\label{fig:FM2}
\end{figure*}

As for the gluon condensate terms $\langle g_{s}^{2}G^{2}\rangle$ and $\langle f^{3}G^{3} \rangle$, the following integral will be encountered,
\begin{eqnarray}
I_{ijk} = \int {{d^4}k\frac{1}{{{{[{{(k + q)}^2} - m_1^2]}^i}{{[{{(k - p')}^2} - m_2^2]}^j}{{({k^2} - m_3^2)}^k}}}}
\end{eqnarray}
This integral can also be calculated by Cutkosky's rules according to the following transformations,
\begin{eqnarray}
\notag
&&I_{ijk} = \frac{1}{{(i - 1)!(j - 1)!(k - 1)!}}\frac{{{\partial ^{i - 1}}}}{{\partial {A^{i - 1}}}}\frac{{{\partial ^{j - 1}}}}{{\partial {B^{j - 1}}}}\frac{{{\partial ^{k - 1}}}}{{\partial {C^{k - 1}}}}\int {{d^4}k} \\
\notag
&& \times \frac{1}{{[{{(k + q)}^2} - A][{{(k - p')}^2} - B]({k^2} - C)}}{|_{A \to {m_{1,}}B \to {m_{2,}}C \to {m_3}}}\\
\notag
&& \to \frac{{{{( - 2\pi i)}^3}}}{{{{(2\pi i)}^2}}}\frac{1}{{(i - 1)!(j - 1)!(k - 1)!}}\frac{{{\partial ^{i - 1}}}}{{\partial {A^{i - 1}}}}\frac{{{\partial ^{j - 1}}}}{{\partial {B^{j - 1}}}}\frac{{{\partial ^{k - 1}}}}{{\partial {C^{k - 1}}}} \\
\notag
&& \times \int {{d^4}k} \delta [{(k + q)^2} - A]\delta [{(k - p')^2} - B]\\
\notag
&&\times \delta ({k^2} - C){|_{A \to {m_{1,}}B \to {m_{2,}}C \to {m_3}}} \\
\notag
&& = \frac{{{{( - 2\pi i)}^3}}}{{{{(2\pi i)}^2}}}\frac{1}{{(i - 1)!(j - 1)!(k - 1)!}}\frac{{{\partial ^{i - 1}}}}{{\partial {A^{i - 1}}}}\frac{{{\partial ^{j - 1}}}}{{\partial {B^{j - 1}}}}\frac{{{\partial ^{k - 1}}}}{{\partial {C^{k - 1}}}}\\
&& \times \frac{\pi }{{2\sqrt {\lambda (s,u,{q^2})} }}{|_{A \to {m_{1,}}B \to {m_{2,}}C \to {m_3}}}
\end{eqnarray}
Besides of these above contributions, the condensate terms of $\langle\overline{q}q\rangle$, $\langle\overline{q}g_{s}\sigma Gq\rangle$ and $\langle\overline{q}q\rangle\langle g_{s}^{2}G^{2}\rangle$ are also taken into account in this work. The feynman diagrams of these condensate terms can be classified into two groups(see Fig. \ref{fig:FM1} and Fig. \ref{fig:FM2}) which correspond to the correlation functions $\Pi_{1}$ and $\Pi_{2}$ in Eq. (\ref{eq:11}), respectively.

\subsection{The electromagnetic form factor}\label{sec3.3}

We take the change of variables $p^{2} \to -P^{2}$, $p'^{2} \to -P'^{2}$ and $q^{2} \to -Q^{2}$ and perform double Borel transforms\cite{Ioffe:1982ia,Ioffe:1982qb} for the variables $P^{2}$ and $P'^{2}$ to both phenomenological and QCD sides. And then $P^{2}$ and $P'^{2}$ will be replaced by $T_{1}^{2}$ and $T_{2}^{2}$ which are called Borel parameters. After matching the phenomenological and QCD side by using quark-hadron duality, the sum rule for the electromagnetic form factor is obtained as,
\begin{eqnarray}\label{eq:24}
\notag
V({Q^2}) &&= \frac{{({m_q} + {m_Q})({m_\mathbb P} + {m_\mathbb V})}}{{{f_\mathbb P}m_\mathbb P^2{f_\mathbb V}{m_\mathbb V}}}{e^{m_\mathbb P^2/T_1^2}}{e^{m_\mathbb V^2/T_2^2}}\\
\notag
&&\times \Big \{ {e_Q}\Big[ - \int\limits_{{s_1}}^{{s_0}} {\int\limits_{{u_1}}^{{u_0}} {{\rho _1}(s,u,{Q^2}){e^{ - u/T_1^2}}{e^{ - s/T_2^2}}duds{|_{|{b_1}(s,u,{Q^2}) \le 1|}}} } \\
\notag
&& +\mathscr{BB}[\Pi_{1}^{\langle\overline{q}q\rangle}+\Pi_{1}^{\langle\overline{q}g_{s}\sigma Gq\rangle}+\Pi_{1}^{\langle\overline{q}q\rangle\langle g_{s}^{2}G^{2}\rangle}]\Big]\\
\notag
&& + {e_q}\Big[ - \int\limits_{{s_1}}^{{s_0}} {\int\limits_{{u_1}}^{{u_0}} {{\rho _2}(s,u,{Q^2}){e^{ - u/T_1^2}}{e^{ - s/T_2^2}}duds{|_{|{b_2}(s,u,{Q^2}) \le 1|}}} }\Big ] \Big \} \\
&& = {e_Q}{V_1}({Q^2}) + {e_q}{V_2}({Q^2})
\end{eqnarray}
where,
\begin{eqnarray}
\notag
&&{b_1}(s,u,{Q^2}) =  \\
\notag
&&\frac{{(u + {Q^2} + m_Q^2 - m_q^2)(s + u + {Q^2}) - 2s(u + m_Q^2 - m_q^2)}}{{\sqrt {{{(u + {Q^2} + m_Q^2 - m_q^2)}^2} - 4sm_Q^2} \sqrt {\lambda (s,u,{Q^2})} }}\\
\notag
&&{b_2}(s,u,{Q^2}) =  \\
\notag
&&\frac{{(u + {Q^2} + m_q^2 - m_Q^2)(s + u + {Q^2}) - 2s(u + m_q^2 - m_Q^2)}}{{\sqrt {{{(u + {Q^2} + m_q^2 - m_Q^2)}^2} - 4sm_q^2} \sqrt {\lambda (s,u,{Q^2})} }}  \\
\end{eqnarray}
In Eq. (\ref{eq:24}), $\mathscr{BB}[~]$ denote the double Borel transforms. It needs to be explained that the vacuum condensate terms $\langle\overline{q}q\rangle$, $\langle\overline{q}g_{s}\sigma Gq\rangle$ and $\langle\overline{q}q\rangle\langle g_{s}^{2}G^{2}\rangle$ for the electromagnetic form factor $V_{2}(Q^{2})$ in Eq. (\ref{eq:24}) will vanish after taking double Borel transforms. The threshold parameters $s_{0}$ and $u_{0}$ are introduced in dispersion integral to eliminate the contributions of higher resonances and continuum states. They should satisfy the relations
$m^{2}_{\mathbb V} \le s_{0} \le m'^{2}_{\mathbb V}$ and $m^{2}_{\mathbb P} \le u_{0} \le m'^{2}_{\mathbb P}$, where $m_{\mathbb V[\mathbb P]}$ and $m'_{\mathbb V[\mathbb P]}$ are the masses of the ground and the first excited state of the vector (pseudoscalar) heavy-light mesons. The masses of the ground and excited states commonly fulfill the relation $m'_{\mathbb V[\mathbb P]}=m_{\mathbb V[\mathbb P]}+\Delta$, where $\Delta$ is usually taken as the value of $0.4 \sim 0.6$ GeV\cite{Bracco:2011pg}.

\section{Numerical results and Discussions}\label{sec4}

The masses of the heavy and light quarks, and the values of vacuum condensate terms are energy-scale dependent, which can be expressed as the following forms according to the re-normalization group equation(RGE),
\begin{eqnarray}
	\notag
	{m_b}(\mu ) &&= {m_b}({m_b}){[\frac{{{\alpha _s}(\mu )}}{{{\alpha _s}({m_b})}}]^{\frac{{12}}{{33-2n_{f}}}}}\\
	\notag
	{m_c}(\mu ) &&= {m_c}({m_c}){[\frac{{{\alpha _s}(\mu )}}{{{\alpha _s}({m_c})}}]^{\frac{{12}}{{33-2n_{f}}}}}\\
	\notag
	{m_s}(\mu ) &&= {m_s}(2\mathrm{GeV}){[\frac{{{\alpha _s}(\mu )}}{{{\alpha _s}(2\mathrm{GeV})}}]^{\frac{12}{33-2n_{f}}}}\\
	\notag
	\left\langle {\bar qq} \right\rangle (\mu ) &&= \left\langle {\bar qq} \right\rangle (1\mathrm{GeV}){[\frac{{{\alpha _s}(1\mathrm{GeV})}}{{{\alpha _s}(\mu )}}]^{\frac{12}{33-2n_{f}}}}\\
	\notag
\end{eqnarray}
\begin{eqnarray}
	\notag
	\left\langle {\bar ss} \right\rangle (\mu ) &&= \left\langle {\bar ss} \right\rangle (1\mathrm{GeV}){[\frac{{{\alpha _s}(1\mathrm{GeV})}}{{{\alpha _s}(\mu )}}]^{\frac{12}{33-2n_{f}}}}\\
	\notag
	\left\langle {\bar q{g_s}\sigma Gq} \right\rangle (\mu ) &&= \left\langle {\bar q{g_s}\sigma Gq} \right\rangle (1\mathrm{GeV}){[\frac{{{\alpha _s}(1\mathrm{GeV})}}{{{\alpha _s}(\mu )}}]^{\frac{2}{{33-2n_{f}}}}}\\
	\notag
	\left\langle {\bar s{g_s}\sigma Gs} \right\rangle (\mu ) &&= \left\langle {\bar s{g_s}\sigma Gs} \right\rangle (1\mathrm{GeV}){[\frac{{{\alpha _s}(1\mathrm{GeV})}}{{{\alpha _s}(\mu )}}]^{\frac{2}{{33-2n_{f}}}}}\\
	\notag
	{\alpha _s}(\mu ) &&= \frac{1}{{{b_0}t}}[1 - \frac{{{b_1}}}{{b_0^2}}\frac{{\log t}}{t}\\
	&& + \frac{{b_1^2({{\log }^2}t - \log t - 1) + {b_0}{b_2}}}{{b_0^4{t^2}}}]
\end{eqnarray}
where $t=\mathrm{log}\frac{\mu^{2}}{\Lambda_{QCD}^{2}}$, $b_{0}=\frac{33-2n_{f}}{12\pi}$, $b_{1}=\frac{153-19n_{f}}{24\pi^{2}}$, $b_{2}=\frac{2857-\frac{5033}{9}n_{f}+\frac{325}{27}n_{f}^{2}}{128\pi^{3}}$, $\Lambda_{QCD}=210$ MeV, 292 MeV and 332 MeV for the flavors $n_{f}=$5, 4 and 3, respectively\cite{ParticleDataGroup:2022pth}.
The $\overline{\mathrm{MS}}$ masses of the heavy and light quarks are adopted from the PDG\cite{ParticleDataGroup:2022pth}, where $m_{c}(m_{c})=1.275\pm0.025$ GeV, $m_{b}(m_{b})=4.18\pm0.03$ GeV, $m_{u(d)}(\mu=1 \mathrm{GeV})=0.006\pm0.001$ GeV and $m_{s}(\mu=2 \mathrm{GeV})=0.095\pm0.005$ GeV. As for the values of other input parameters, they are all listed in Table \ref{HP}, where the quark and quark-gluon condensate parameters take their values in the energy scale $\mu=1$ GeV.
Based on our previous researches about the heavy-light mesons\cite{Wang:2015mxa}, the values of decay constants are selected by following energy-scale, where $\mu=1$ GeV for $D^{*}\to D$ electromagnetic form factor, $\mu=2.5$ GeV for $B^{*}\to B$, $\mu=1.1$ GeV for $D^{*}_{s}\to D_{s}$ and $\mu=2.6$ GeV for $B^{*}_{s}\to B_{s}$.

\begin{table}[htbp]
	\begin{ruledtabular}\caption{Input parameters (IP) in this work.}
		\label{HP}
		\begin{tabular}{c c c c}
			IP&values(GeV)&IP&values\\ \hline
			$m_{D}$&1.86\cite{ParticleDataGroup:2022pth}&$f_{B}$&$0.192\pm0.013$ GeV\cite{Wang:2015mxa}\\
			$m_{D^{*}}$&2.01\cite{ParticleDataGroup:2022pth}&$f_{B^{*}}$&$0.213\pm0.018$ GeV\cite{Wang:2015mxa}\\
			$m_{D_{s}}$&1.97\cite{ParticleDataGroup:2022pth}&$f_{B_{s}}$&$0.230\pm0.013$ GeV\cite{Wang:2015mxa}\\
			$m_{D_{s}^{*}}$&2.11\cite{ParticleDataGroup:2022pth}&$f_{B^{*}_{s}}$&$0.255\pm0.019$ GeV\cite{Wang:2015mxa}\\
			$m_{B}$&5.28\cite{ParticleDataGroup:2022pth}&$\langle\overline{q}q\rangle$&$-(0.23\pm0.01)^{3}$ GeV$^{3}$\cite{Shifman:1978by,Reinders:1984sr}\\
			$m_{B^{*}}$&5.32\cite{ParticleDataGroup:2022pth}&$\langle\overline{q}g_{s}\sigma Gq\rangle$&$m_{0}^{2}\langle\overline{q}q\rangle$\cite{Shifman:1978by,Reinders:1984sr} \\
			$m_{B_{s}}$&5.37\cite{ParticleDataGroup:2022pth}&$\langle\overline{s}s\rangle$&$-(0.8\pm0.1)\langle\overline{q}q\rangle$\cite{Shifman:1978by,Reinders:1984sr}\\
			$m_{B^{*}_{s}}$&5.42\cite{ParticleDataGroup:2022pth}&$\langle\overline{s}g_{s}\sigma Gs\rangle$&$m_{0}^{2}\langle\overline{s}s\rangle$\cite{Shifman:1978by,Reinders:1984sr}\\
			$f_{D}$&$0.210\pm0.011$ \cite{Wang:2015mxa}&$m_{0}^{2}$&$0.8\pm0.1$ GeV$^{2}$\cite{Shifman:1978by,Reinders:1984sr}\\
			$f_{D^{*}}$&$0.236\pm0.021$ \cite{Wang:2015mxa}&$\langle g_{s}^{2}G^{2}\rangle$&$0.88\pm0.15$ GeV$^{4}$\cite{Narison:2010cg,Narison:2011xe,Narison:2011rn}\\
			$f_{D_{s}}$&$0.259\pm0.010$\cite{Wang:2015mxa}&$\langle f^{3}G^{3}\rangle$&$(8.8\pm5.5)\langle\alpha_{s}G^{2}\rangle$ \cite{Narison:2010cg,Narison:2011xe,Narison:2011rn}\\
			$f_{D_{s}^{*}}$&$0.308\pm0.021$\cite{Wang:2015mxa}&~&~\\
		\end{tabular}
	\end{ruledtabular}
\end{table}

In the framework of QCDSR, two criteria should be satisfied, which are the pole dominance and convergence of operator product expansion (OPE). The pole and continuum contribution can be defined as\cite{Bracco:2011pg},
\begin{eqnarray}\label{eq:27}
\notag
\mathrm{Pole}=\frac{\Pi^{\mathrm{OPE}}_{\mathrm{pole}}(T_{1}^{2},T_{2}^{2})}{\Pi^{\mathrm{OPE}}_{\mathrm{pole}}(T_{1}^{2},T_{2}^{2})+\Pi^{\mathrm{OPE}}_{\mathrm{cont}}(T_{1}^{2},T_{2}^{2})} \\
\mathrm{Continuum}=\frac{\Pi^{\mathrm{OPE}}_{\mathrm{cont}}(T_{1}^{2},T_{2}^{2})}{\Pi^{\mathrm{OPE}}_{\mathrm{pole}}(T_{1}^{2},T_{2}^{2})+\Pi^{\mathrm{OPE}}_{\mathrm{cont}}(T_{1}^{2},T_{2}^{2})}
\end{eqnarray}
with
\begin{eqnarray}\label{eq:28}
\notag
\Pi^{\mathrm{OPE}}_{\mathrm{pole}}(T_{1}^{2},T_{2}^{2})=-\int_{s_{1}}^{s_{0}}\int_{u_{1}}^{u_{0}}\rho^{\mathrm{OPE}}(s,u,Q^2)e^{-\frac{s}{T_{1}^{2}}}e^{-\frac{u}{T_{2}^{2}}}dsdu \\
\Pi^{\mathrm{OPE}}_{\mathrm{cont}}(T_{1}^{2},T_{2}^{2})=-\int_{s_{0}}^{\infty}\int_{u_{0}}^{\infty}\rho^{\mathrm{OPE}}(s,u,Q^2)e^{-\frac{s}{T_{1}^{2}}}e^{-\frac{u}{T_{2}^{2}}}dsdu
\end{eqnarray}

\begin{figure}[htbp]
\centering
\subfigure[]{\includegraphics[width=4cm]{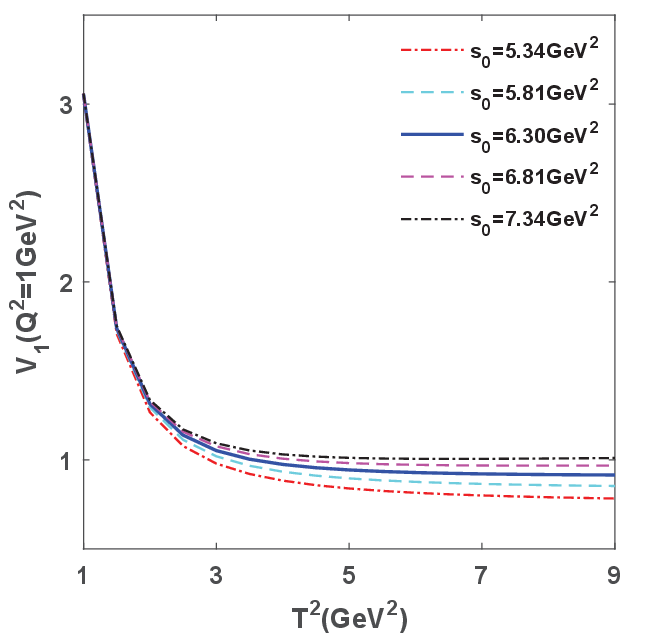}}
\subfigure[]{\includegraphics[width=4cm]{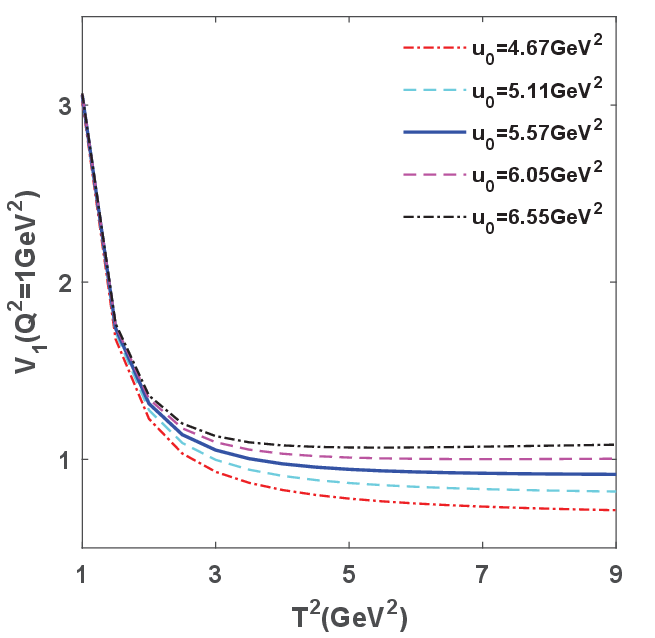}}
\caption{The $D^{*}\to D$ electromagnetic form factor $V_{1}$ on Borel parameter $T^{2}$ with different $s_{0}$(a) and $u_{0}$(b).}
\label{s0u0}
\end{figure}

It can be seen from Eqs. (\ref{eq:24}) and (\ref{eq:27}) that the results of QCDSR depend on input parameters such as the Borel parameters $T_{1}^{2}$ and $T_{2}^{2}$, the continuum threshold parameters $s_{0}$ and $u_{0}$, and the square momentum $Q^{2}$. In this article, we take $T_{1}^{2}=T^{2}$ and $T_{2}^{2}=kT_{1}^{2}=kT^{2}$. $k$ is a constant which is related to the ratio of vector and pseudoscalar heavy-light mesons and can be expressed as $k=\frac{m^{2}_{\mathbb{V}}}{m^{2}_{\mathbb{P}}}$. The threshold parameters are defined as $s_{0}=(m_{\mathbb{V}}+\Delta_{\mathbb{V}})^{2}$ and $u_{0}=(m_{\mathbb{P}}+\Delta_{\mathbb{P}})^{2}$. The values of $\Delta_{\mathbb{V}}$ and $\Delta_{\mathbb{P}}$ should be smaller than the experimental value of the distance between the ground and first excited state. Taking the electromagnetic form factor $V_{1}$ of $D^{*}\to D$ as an example, we simply discuss how to select $s_{0}$ and $u_{0}$. Fixing $Q^{2}=1$GeV$^{2}$ in Eq. (\ref{eq:28}), we firstly plot the electromagnetic form factor $V_{1}$ on Borel parameter $T^{2}$ in Fig. \ref{s0u0}, where different values of $s_{0}$ and $u_{0}$ are adopted. It is indicated that the results have good stability when $s_{0}$ is taken to be $6.30$ GeV$^{2}$($\Delta_{\mathbb{V}}$=0.5 GeV) and $u_{0}$ is $5.57$ GeV$^{2}$($\Delta_{\mathbb{P}}$=0.5 GeV). Thus, the final results will be obtained by taking $\Delta_{\mathbb{V}}=\Delta_{\mathbb{P}}=0.4$, 0.5 and 0.6 GeV, where 0.5 GeV is used to determine the central values of the electromagnetic form factors, and 0.4, 0.6 GeV are for the lower and upper bounds of the final results.

An appropriate region of the Borel parameter $T^{2}$ needs to be chosen to make the final results stable and reliable. This region is commonly called 'Borel platform'. After repeated trial and contrast, we finally determine the Borel platforms which are shown in Fig. \ref{BW} in Appendix. From Fig. \ref{BW}, we can see that the values of the form factors have good stability in the Borel regions, which means the convergence of OPE is well satisfied.
In addition, the pole and continuum contributions in the Borel platforms are plotted in Fig. \ref{PC}. We can see from Fig. \ref{PC} that the central values of pole contribution in the Borel platform satisfy the condition of pole dominance ($>50\%$). That is to say, two criteria of pole dominance and convergence of OPE are both satisfied. The Borel platform, pole contributions in the Borel platform, and the values of electromagnetic form factors in $Q^{2}=1$ GeV$^{2}$ are listed in Table~\ref{BPV}.
\begin{table}[htbp]
\begin{ruledtabular}\caption{The Borel platform, pole contributions (Pole) and the values of electromagnetic form-factors in $Q^{2}=1$ GeV$^{2}$ for different radiative decay modes.}
\label{BPV}
\begin{tabular}{c c c c c}
Mode&Form factor&Borel platform&Pole(\%)&$V_{i}(Q^{2}=1$ GeV$^{2})$ \\ \hline
\multirow{2}*{$D^{*}\to D \gamma$}&$V_{1}$ &$3\sim5$&$60\sim44$&$0.97^{+0.10}_{-0.09}$ \\
~&$V_{2}$ &$3.5\sim5.5$&$68\sim44$&$0.69^{+0.17}_{-0.16}$  \\  \hline
\multirow{2}*{$D_{s}^{*}\to D_{s} \gamma$}&$V_{1}$ &$4.5\sim6.5$&$58\sim42$&$0.98^{+0.12}_{-0.11}$ \\
~&$V_{2}$ &$5\sim7$&$64\sim46$&$0.95^{+0.20}_{-0.17}$  \\  \hline
\multirow{2}*{$B^{*}\to B \gamma$}&$V_{1}$ &$11\sim13$&$53\sim46$&$1.42^{+0.13}_{-0.12}$ \\
~&$V_{2}$ &$12\sim14$&$58\sim49$&$2.54^{+0.46}_{-0.47}$  \\  \hline
\multirow{2}*{$B_{s}^{*}\to B_{s} \gamma$}&$V_{1}$ &$13\sim15$&$52\sim45$&$1.30^{+0.12}_{-0.13}$ \\
~&$V_{2}$ &$15\sim17$&$57\sim50$&$3.00^{+0.45}_{-0.48}$  \\
\end{tabular}
\end{ruledtabular}
\end{table}

\begin{figure}[htbp]
	\centering
	\subfigure[]{\includegraphics[width=4cm]{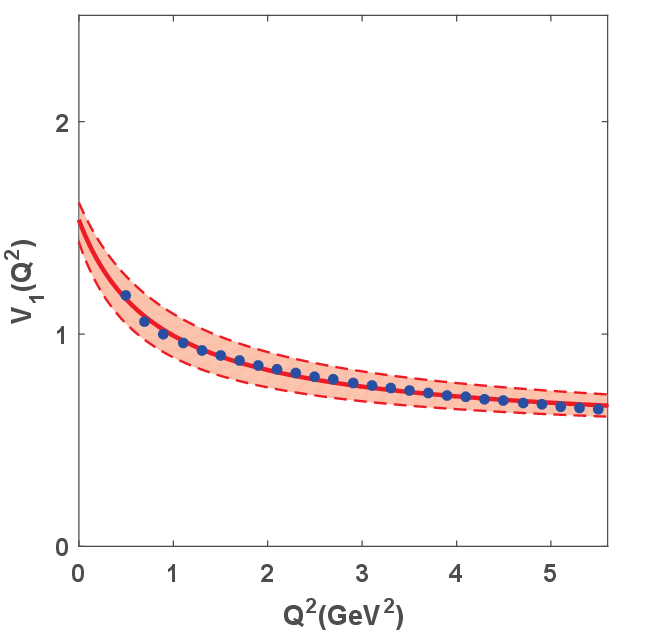}}
	\subfigure[]{\includegraphics[width=4cm]{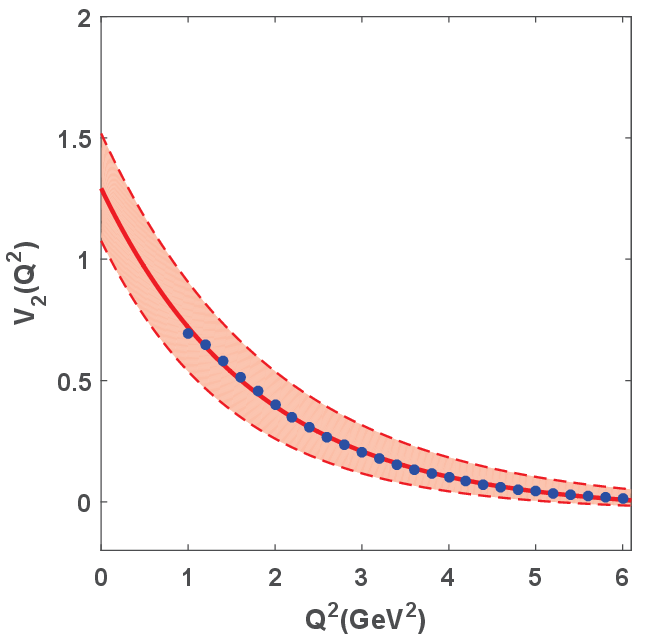}}
	
	\subfigure[]{\includegraphics[width=4cm]{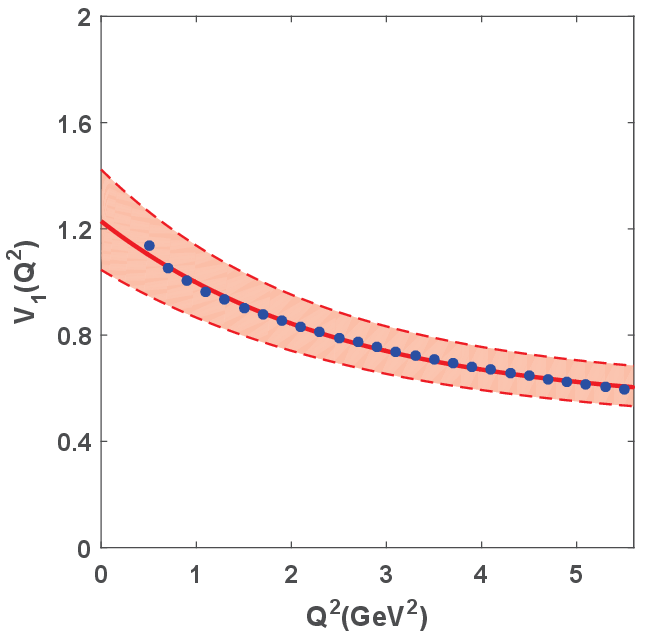}}
	\subfigure[]{\includegraphics[width=4cm]{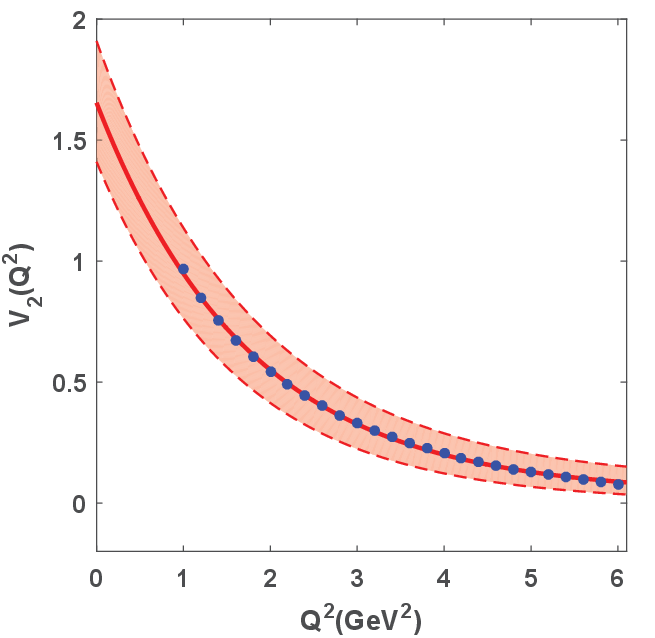}}
	\caption{The fitting results for $D^{*0(\pm)} \to D^{0(\pm)}$ (a,b) and $D_{s}^{*\pm} \to D_{s}^{\pm}$ (c,d). The blue circles are the form factors in space-like regions($Q^2>0$), the red solid lines are the fitting curves, and the red bands represent the error regions.}
	\label{FF1}
\end{figure}

\begin{figure}[htbp]
	\centering
	
	\subfigure[]{\includegraphics[width=4cm]{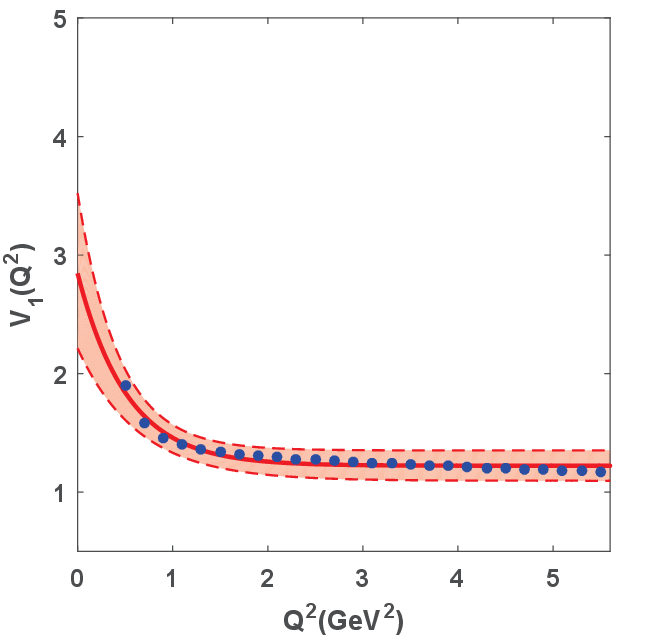}}
	\subfigure[]{\includegraphics[width=4cm]{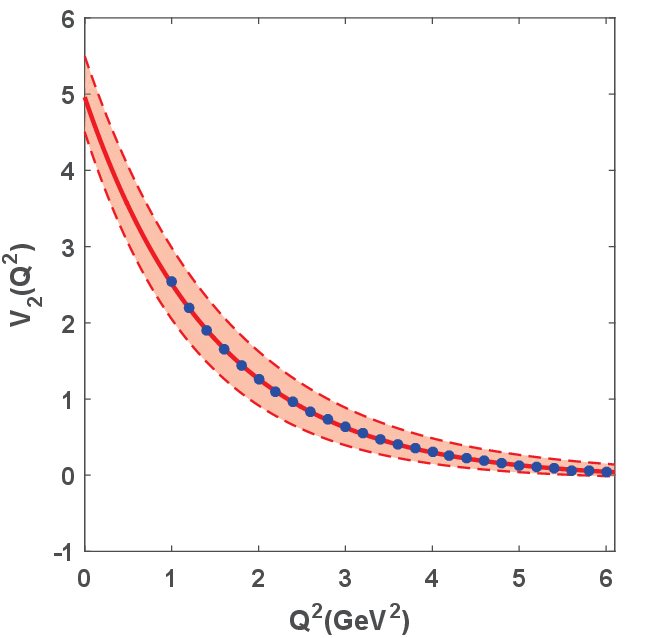}}
	
	\subfigure[]{\includegraphics[width=4cm]{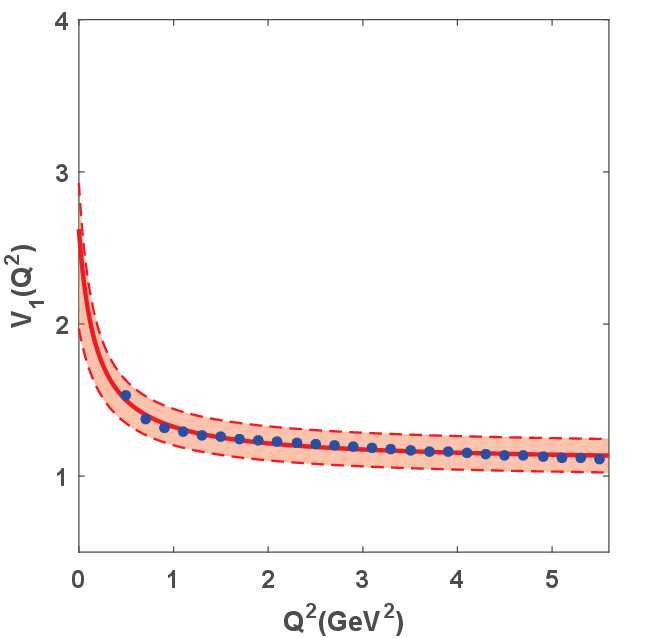}}
	\subfigure[]{\includegraphics[width=4cm]{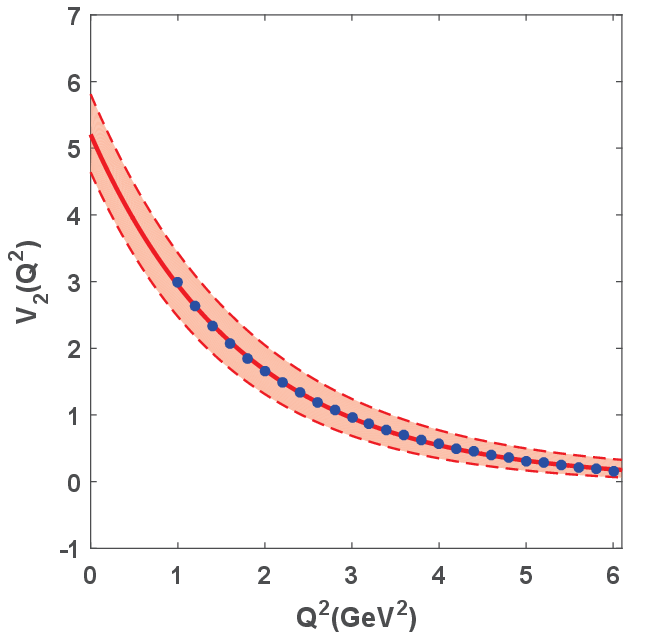}}
	\caption{The fitting results for $B^{*0(\pm)} \to B^{0(\pm)}$ (a,b) and $B_{s}^{*0} \to B_{s}^{0}$ (c,d). The blue circles are the form factors in space-like regions($Q^2>0$), the red solid lines are the fitting curves, and the red bands represent the error regions.}
	\label{FF2}
\end{figure}

By taking different values of $Q^{2}$, the electromagnetic form factors $V_{i}(Q^{2})$ in space-like regions ($Q^2>0$) can be obtained, where $Q^{2}$ is in the range of $0.5\sim5.5$ GeV$^{2}$ for $V_{1}(Q^{2})$, and $1\sim6$ GeV$^{2}$ for $V_{2}(Q^{2})$. For the values of the electromagnetic form factors at $Q^{2}=0$, which are used to analyze the radiative decay of the vector heavy-light mesons, they can be obtained by extrapolating the results $V_{i}(Q^{2})$ into $Q^{2}=0$. This process is realized by selecting appropriate analytical functions to fit the results $V_{i}(Q^{2})$ in space-like region. In general, the single pole fitting or exponential fitting are usually used to roughly parameterize the electroweak form factors. More precise and physical method is the $z-$ series expansion which is used to extrapolate the heavy-to-light electroweak form factors to large momentum transfer regions$(q^2=-Q^2>>0)$\cite{Boyd:1994tt,Wang:2015vgv,Cui:2022zwm}. For the electromagnetic form factors of vector heavy-light mesons to pseudoscalar heavy-light mesons, both the initial and final states are heavy mesons, and we only take care of the values at $Q^2=0$. Thus, we do not directly employ the $z-$ series expansion method to fit the electromagnetic form factors. After repeated trial and error, we find that the results in space-like regions ($Q^2>0$) can be well fitted by the combinations of the single pole, exponential and polynomial functions. In addition, we know that the series expansion of the exponential function and $z-$ series expansion are essentially consistent with each other in their mathematical form. Thus, the simple parameterizations method is adopted to fit the form factors in this work, and it has the following form,
\begin{eqnarray}\label{eq:29}
V_{i}(Q^{2})=\frac{A}{1+BQ^{2}}Exp(-CQ^{2})+D
\end{eqnarray}
where $A$, $B$, $C$ and $D$ are the fitting parameters and their values are listed in Table~\ref{FFSC}. The fitting results are also explicitly shown in Figs. \ref{FF1} and \ref{FF2}. By setting $Q^{2}=0$ in the analytical functions(Eq. (\ref{eq:29})), the values of electromagnetic form factors in $Q^{2}=0$($V_{i}(0)$) are obtained, which are also listed in the last column of Table~\ref{FFSC}.

\begin{table}[htbp]
\begin{ruledtabular}\caption{The parameters for the analysis function and the values of electromagnetic form factors in $Q^{2}=0$ for different radiative decay modes.}
\label{FFSC}
\begin{tabular}{c c c c c c c}
Mode&Form factor&$A$&$B$&$C$&$D$&$V_{i}(0)$ \\ \hline
\multirow{2}*{$D^{*}\to D \gamma$}&$V_{1}$ &$1.01$&$1.18$&$-0.0006$&$0.53$&$1.54^{+0.08}_{-0.11}$ \\
~&$V_{2}$ &$1.33$&$0$&$0.56$&$-0.04$&$1.29^{+0.23}_{-0.21}$  \\  \hline
\multirow{2}*{$D_{s}^{*}\to D_{s} \gamma$}&$V_{1}$ &$0.70$&$0$&$0.40$&$0.53$&$1.23^{+0.19}_{-0.18}$ \\
~&$V_{2}$ &$1.62$&$0$&$0.57$&$0.04$&$1.65^{+0.26}_{-0.24}$  \\  \hline
\multirow{2}*{$B^{*}\to B \gamma$}&$V_{1}$ &$1.62$&$0$&$1.93$&$1.22$&$2.85^{+0.67}_{-0.64}$ \\
~&$V_{2}$ &$5.00$&$0$&$0.67$&$-0.04$&$4.96^{+0.54}_{-0.46}$  \\  \hline
\multirow{2}*{$B_{s}^{*}\to B_{s} \gamma$}&$V_{1}$ &$1.54$&$5.46$&$0.0002$&$1.09$&$2.62^{+0.31}_{-0.65}$ \\
~&$V_{2}$ &$5.19$&$0$&$0.57$&$0.01$&$5.20^{+0.61}_{-0.56}$  \\
\end{tabular}
\end{ruledtabular}
\end{table}

According to the heavy-hadron chiral perturbation theory, the relation of the form factors $\frac{V_{1}(0)}{V_{2}(0)}<1$ should be satisfied\cite{Cheng:1992xi}. From Table~\ref{FFSC}, we find the values of $\frac{V_{1}(0)}{V_{2}(0)}$ for $D_{s}^{*}\to D_{s}$, $B^{*}\to B$ and $B_{s}^{*}\to B_{s}$ are all lower than 1, which means the results are consistent well with the conclusion of heavy-hadron chiral perturbation theory. For the form factor of $D^{*}\to D$, the central value of $V_{1}(0)$ is slightly larger than that of $V_{2}(0)$, which looks as if these results are contradict with the conclusion of chiral perturbation theory. Actually, if we consider the error bar of the results, the relation $\frac{V_{1}(0)}{V_{2}(0)}<1$ is still roughly satisfied. It is known that the next-to-leading order contribution is important in small $Q^{2}$ regions. However, the calculation of this diagram is very complicated for three-point QCDSR, thus we do not consider this contribution, which may lead to errors in the final results. In addition, the parametrization of the form factors is model dependent, it may also lead to some errors in the results. According to these above discussions, we think there are no contradiction between the results of QCDSR and that of the heavy-hadron chiral perturbation theory.    

Basing on these above analysis about the electromagnetic form factors, we can now discuss the radiative decays of vector heavy-light mesons. The standard form of two body decay width can be written as,
\begin{eqnarray}\label{eq:30}
\notag
\Gamma  &&= \frac{1}{{2J + 1}}\sum {\frac{p}{{8\pi M_i^2}}|T{|^2}} \\
p &&= \frac{{\sqrt {[M_i^2 - {{(m + {M_f})}^2}][M_i^2 - {{({M_f} - m)}^2}]} }}{{2{M_i}}}
\end{eqnarray}
where $M_{i}$ and $M_{f}$ represent the masses of initial and final mesons, $J$ is the total angular momentum of the initial meson, $\Sigma$ denotes the summation of all the polarization vectors, and $T$ is the scattering amplitude.

With Eqs. (\ref{eq:4}) and (\ref{eq:30}), the decay width of vector heavy-light meson to pseudoscalar heavy-light meson and photon can be expressed as,
\begin{eqnarray}
\Gamma= \frac{{\alpha |Q_Q V_1(0)+Q_q V_2(0){|^2}({m_\mathbb V} + {m_\mathbb P}){{({m_\mathbb V} - {m_\mathbb P})}^3}}}{{24m_\mathbb V^3}}
\end{eqnarray}
Here, $Q_{Q[q]}$ denote the electric charges of quarks, which are taken as $\frac{2}{3}$ for u and c quarks, $-\frac{1}{3}$ for d, s and b quarks. $\alpha$ is the electromagnetic fine structure constant whose value is $\frac{1}{137}$. Using the values about electromagnetic form factors $V_{i}(0)$ in Table~\ref{FFSC}, we can obtain the decay widths of radiative decay for vector heavy-light mesons. The results of present work and those of other collaboration's are all listed in Table \ref{FS}.

According to Particle Data Group(PDG)\cite{ParticleDataGroup:2022pth}, the experimental data of the total decay width for $D^{*+}$ is $83.4\pm1.8$ keV, and the ratio of the decay width of $D^{*+}\to D^{+}\gamma$ to the total width is about $(1.6\pm0.4)\%$. The results of present work about this radiative decay is $0.17^{+0.08}_{-0.07}$ keV, which is about 0.20\% of the total width $83.4\pm1.8$ keV and is smaller than the experimental data. In Ref\cite{Aliev:1994nq}, the radiative decay for $D^{*+}$ was also analyzed by QCDSR, and their result is compatible with ours, which is also smaller than the experimental data. The total widths for $D^{*0}$ and $D_{s}^{*}$ are given as $\Gamma(D^{*0})<2.1$ MeV and $\Gamma(D_{s}^{*})<1.9$ MeV, and the ratios of the radiative decays are about $(35.3\pm0.9)$\% and $(93.5\pm0.7)$\% of the total widths. However, the exact values of these decay widths have not be determined yet. Our results for $D^{*0}$ and $D_{s}^{*}$ are $1.74^{+0.40}_{-0.37}$ keV and $0.029^{+0.009}_{-0.008}$ keV, which are compatible with the experiment data. In addition, the electromagnetic form factor which has the same expression as ours was firstly analyzed by Lattice QCD\cite{Donald:2013sra}, their radiative decay width about $D_{s}^{*}\to D_{s}\gamma$ is given as $0.066\pm0.026$ keV. Considering the uncertainties, our result is in agreement well with theirs. We also notice that the radiative decays of the heavy-light mesons were analyzed in the framework of light-cone QCD sum rules(LCSR) by consedering the subleading power corrections in recent years\cite{Li:2020rcg,Pullin:2021ebn}. In Ref\cite{Li:2020rcg}, the electromagnetic coupling constants of the heavy-light mesons were obtained by considering the contributions from twist-two photon distribution amplitude at next-to-leading-order in $\alpha_{s}$. For the convenience of comparing the results, we use their electromagnetic coupling constants to get the corresponding decay widths which are shown in Table. \ref{FS}. As for the results of mesons $B^{*\pm(0)}$ and $B_{s}^{*}$, it can be seen from Table \ref{FS} that the theoretical results of different collaborations are not consistent well with each other, which needs to be further testified by experiments in the future.

\begin{large}
\section{Conclusions}\label{sec5}
\end{large}
In this article, the electromagnetic form factors in space-like regions ($Q^{2}>0$) for vector heavy-light meson to pseudoscalar heavy-light meson are firstly analyzed within the framework of three-point QCDSR. Then, the electromagnetic form factors in $Q^{2}=0$ are obtained by fitting the results into analytical functions in space-like regions. Based on these results, the radiative decays of the vector heavy-light mesons are systematically analyzed. We hope the results about electromagnetic form factors and radiative decays can help to shed more light on the nature of the hadrons.

\section*{Acknowledgements}

This project is supported by National Natural Science Foundation, Grant Number 12175068 and Natural Science Foundation of HeBei Province, Grant Number A2018502124.

\begin{widetext}
\begin{table*}[htbp]
\begin{ruledtabular}\caption{The radiation decay width of vector heavy-light mesons, all values are in units of keV.}
\label{FS}
\begin{tabular}{c c c c c c c c}
Method&Ref&$\Gamma(D^{*0}\to D^{0}\gamma)$&$\Gamma(D^{*+}\to D^{+}\gamma)$&$\Gamma(D_{s}^{*}\to D_{s}\gamma)$&$\Gamma(B^{*0}\to B^{0}\gamma)$&$\Gamma(B^{*+}\to B^{+}\gamma)$&$\Gamma(B_{s}^{*}\to B_{s}\gamma)$ \\ \hline
\multirow{2}*{QCDSR}&present work&$1.74^{+0.40}_{-0.37}$&$0.17^{+0.08}_{-0.07}$&$0.029^{+0.009}_{-0.008}$&$0.018^{+0.006}_{-0.005}$&$0.015^{+0.007}_{-0.007}$&$0.016^{+0.003}_{-0.005}$ \\
~&\cite{Aliev:1994nq}&$2.43\pm0.21$&$0.22\pm0.06$&$0.25\pm0.08$&-&-&- \\ \hline
\multirow{3}*{LCSR}&\cite{Li:2020rcg}&$13.70^{+5.88}_{-4.55}$&$0.14^{+0.27}_{-0.12}$&$0.041^{+0.137}_{-0.041}$&$0.17^{+0.05}_{-0.05}$&$0.42^{+0.14}_{-0.11}$&$0.15^{+0.03}_{-0.04}$ \\
~&\cite{Pullin:2021ebn}&$27.83^{+9.23}_{-9.50}$&$0.96^{+0.58}_{-0.62}$&$2.36^{+1.49}_{-1.41}$&$0.16^{+0.06}_{-0.06}$&$0.45^{+0.17}_{-0.16}$&$0.24^{+0.08}_{-0.08}$ \\
~&\cite{Aliev:1995zlh}&14.40&1.50&-&0.16&0.63&- \\  \hline
Lattice QCD&\cite{Becirevic:2009xp}$^{*}$, \cite{Donald:2013sra}$^{\dagger}$&$27\pm14^{*}$&$0.8\pm0.7^{*}$&$0.066\pm0.026^{\dagger}$&-&-&- \\ \hline
Salpter-like model&\cite{Colangelo:1994jc}&20.8&0.46&0.38&0.092&0.243&0.080 \\ \hline
$\chi$-loop model&\multirow{2}*{\cite{Casalbuoni:1996pg}}&10.3&0.16&0.06&0.09&0.14&0.03 \\
VMD model&~&18.0&0.64&0.35&0.12&0.37&0.09 \\  \hline
\multirow{2}*{Chiral quark model}&\cite{Deandrea:1998uz}&13.11&0.25&-&0.05&-&- \\
~&\cite{Goity:2000dk}&32.02&1.42&0.3&-&-&- \\  \hline
Bag model&\cite{Orsland:1998de}&7.18&1.72&-&0.064&0.272&0.051 \\  \hline
\multirow{2}*{Light front model}&\cite{Choi:2007se}&$20.0\pm0.3$&$0.90\pm0.02$&$0.18\pm0.01$&$0.13\pm0.01$&$0.40\pm0.03$&$0.068\pm0.017$ \\
~&\cite{Jaus:1996np}&21.69&0.56&-&0.142&0.429&- \\  \hline
GI quark model&\cite{Li:2022vby}&97&9.9&-&-&-&- \\  \hline
Covariant confined quark model&\cite{Ivanov:2022nnq}$^{*}$, \cite{Tran:2023hrn}$^{\dagger}$&$18.4\pm7.4^{\dagger}$&$1.21\pm0.48^{\dagger}$&$0.55\pm0.22^{\dagger}$&$0.126\pm0.019^{*}$&$0.372\pm0.056^{*}$&$0.090\pm0.014^{*}$ \\
\end{tabular}
\end{ruledtabular}
\end{table*}
\end{widetext}

\begin{widetext}
	\begin{large}
		\textbf{Appendix: The pole and continuum contributions, and the contributions of different vacuum condensate terms.}
	\end{large}

	\begin{figure*}[htbp]
		\vspace{8pt}
		\centering
		\subfigure[]{\includegraphics[width=4cm]{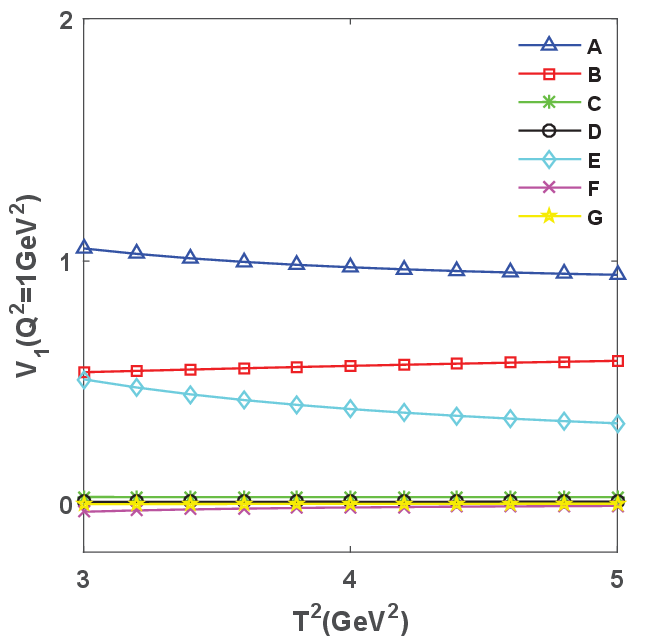}}
		\subfigure[]{\includegraphics[width=4cm]{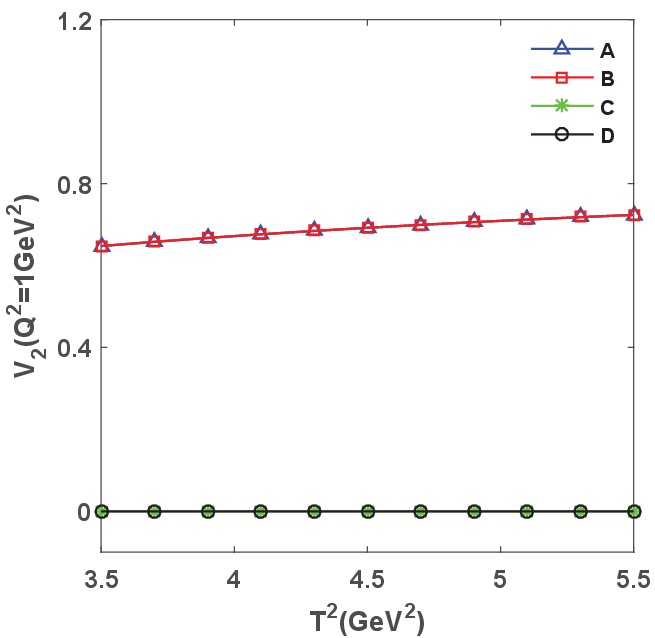}}
		\subfigure[]{\includegraphics[width=4cm]{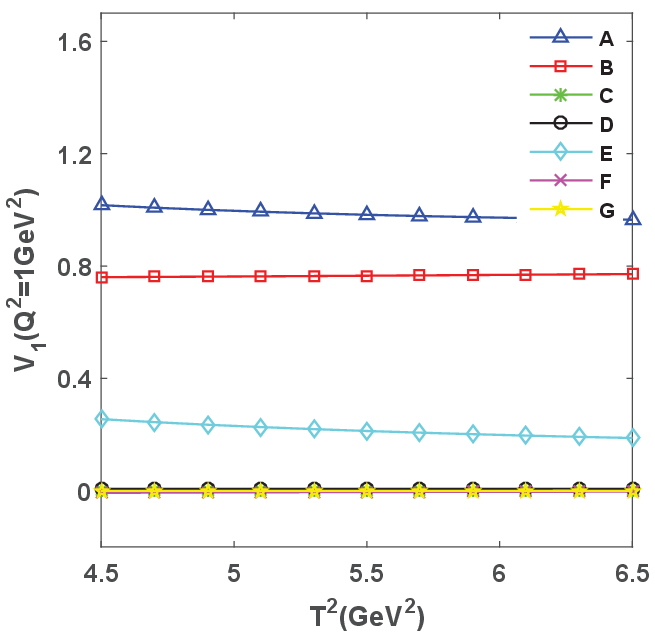}}
		\subfigure[]{\includegraphics[width=4cm]{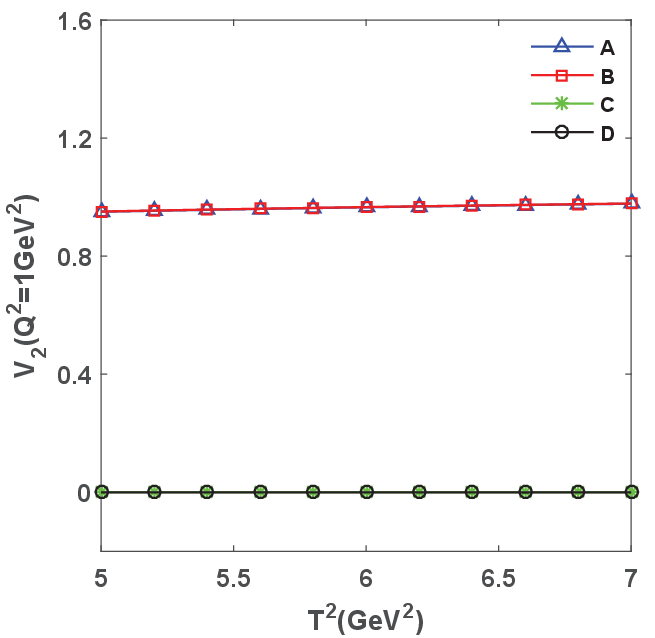}}
		
		\subfigure[]{\includegraphics[width=4cm]{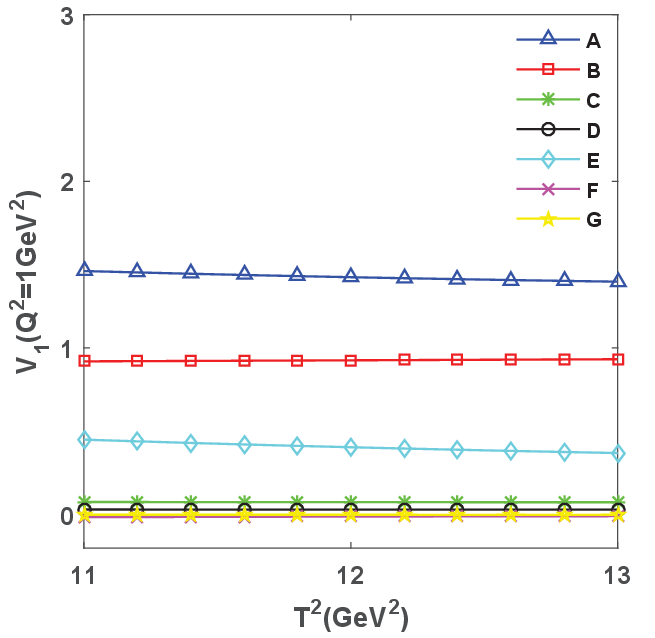}}
		\subfigure[]{\includegraphics[width=4cm]{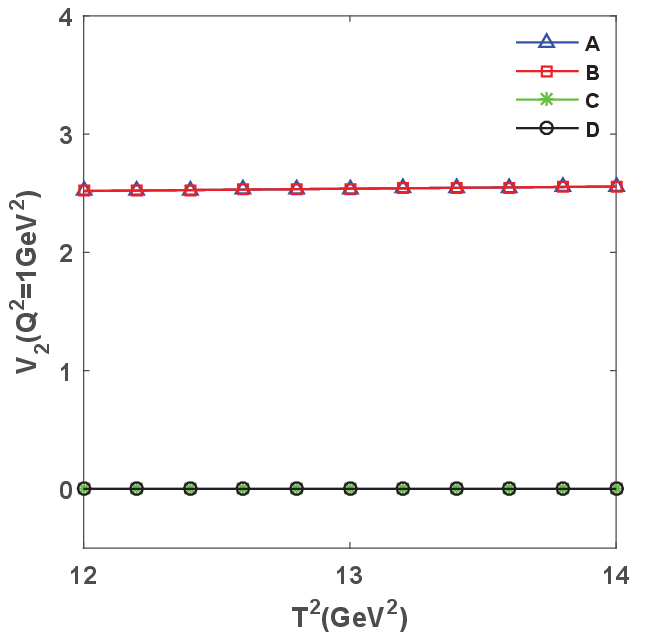}}
		\subfigure[]{\includegraphics[width=4cm]{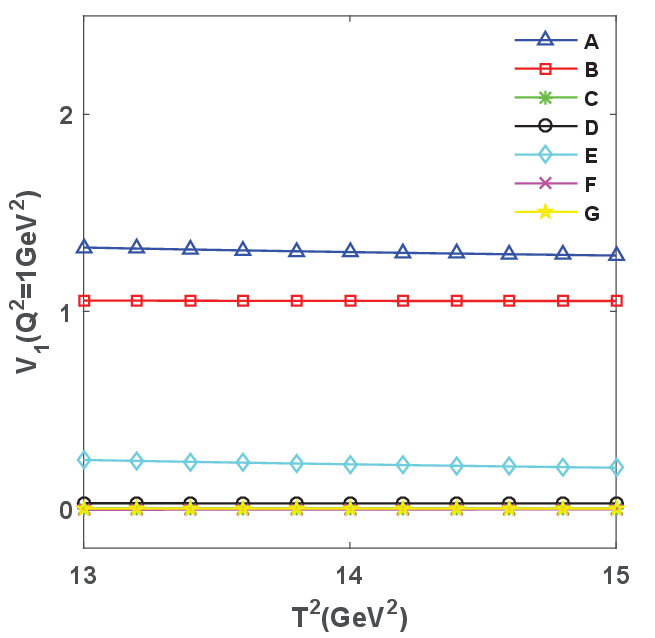}}
		\subfigure[]{\includegraphics[width=4cm]{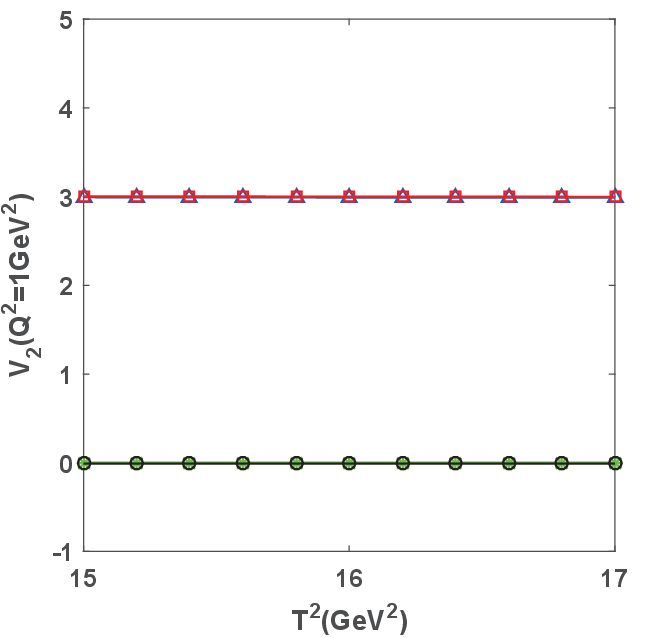}}
		\caption{The contributions of different vacuum condensate terms with variation of the Borel parameter $T^2$ for $D^{*0(\pm)} \to D^{0(\pm)}$ (a,b), $D_{s}^{*\pm} \to D_{s}^{\pm}$ (c,d), $B^{*0(\pm)} \to B^{0(\pm)}$ (e,f), and $B_{s}^{*0} \to B_{s}^{0}$ electromagnetic form-factors, where A-G represent the total, perturbative term, $\langle g_{s}^{2}G^{2}\rangle$, $\langle f^{3} G^{3}\rangle$, $\langle \overline{q}q \rangle$, $\langle \overline{q}g_{s}\sigma Gq\rangle$, and $\langle\overline{q}q \rangle\langle g_{s}^{2}G^{2} \rangle$ contributions, here the $q$ denotes the light quarks $u(d)$ or $s$.}
		\label{BW}
	\end{figure*}
	\begin{figure*}[htbp]
		\vspace{10pt}
		\centering
		\subfigure[]{\includegraphics[width=4cm]{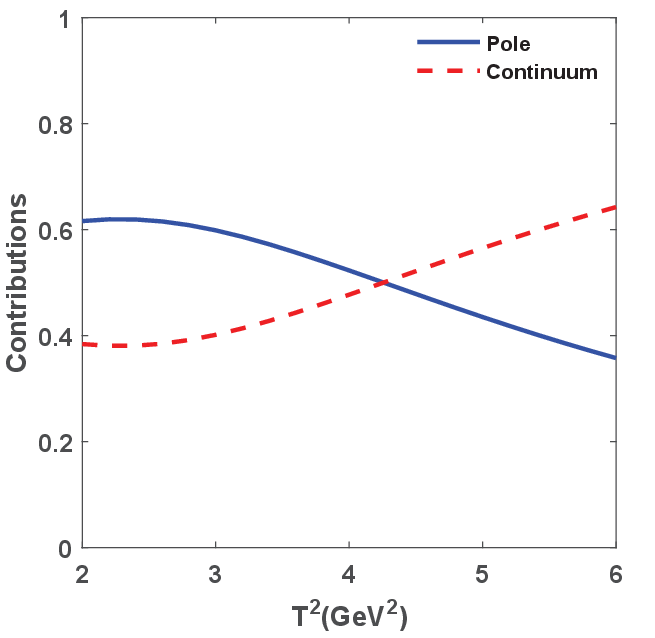}}
		\subfigure[]{\includegraphics[width=4cm]{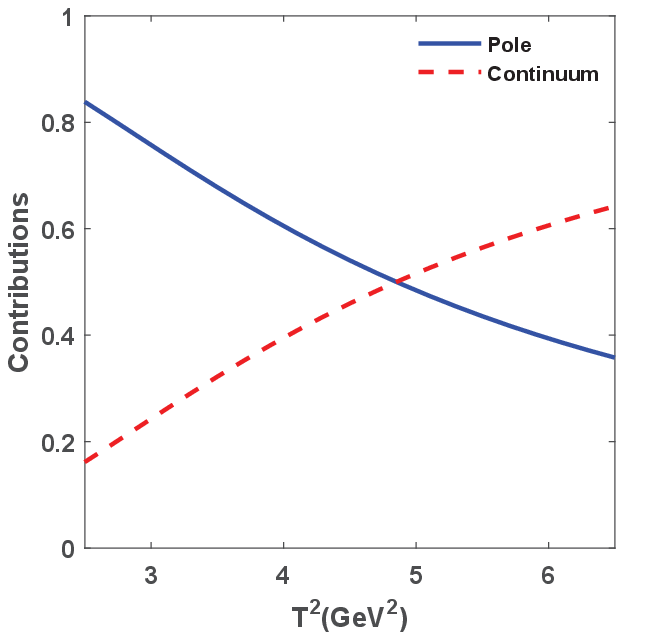}}
		\subfigure[]{\includegraphics[width=4cm]{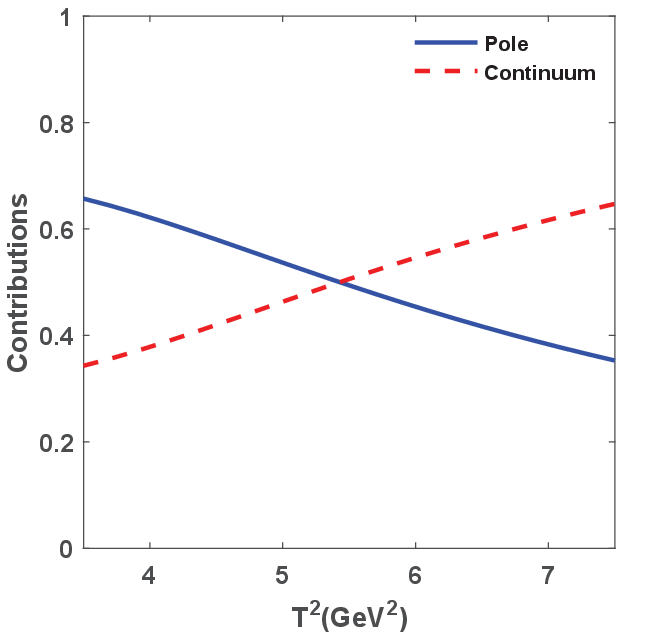}}
		\subfigure[]{\includegraphics[width=4cm]{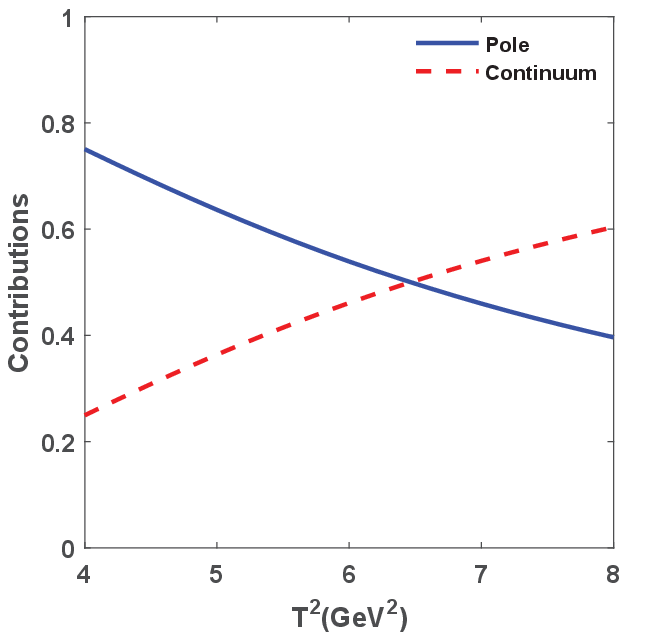}}
		
		\subfigure[]{\includegraphics[width=4cm]{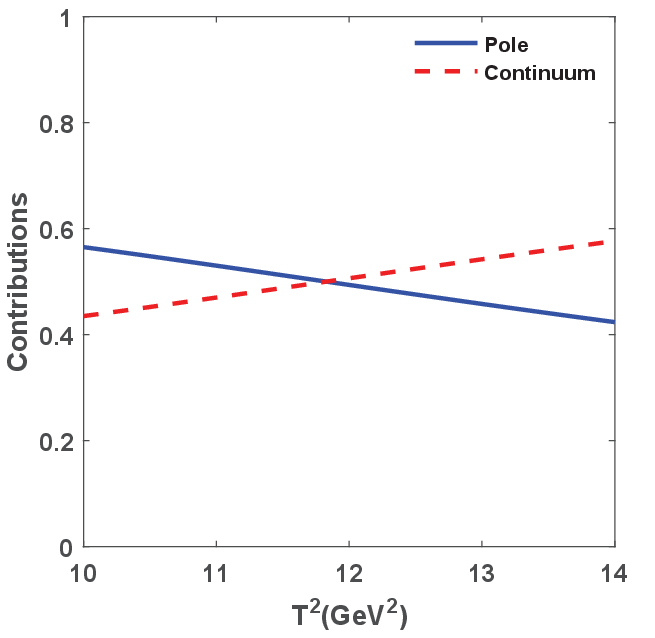}}
		\subfigure[]{\includegraphics[width=4cm]{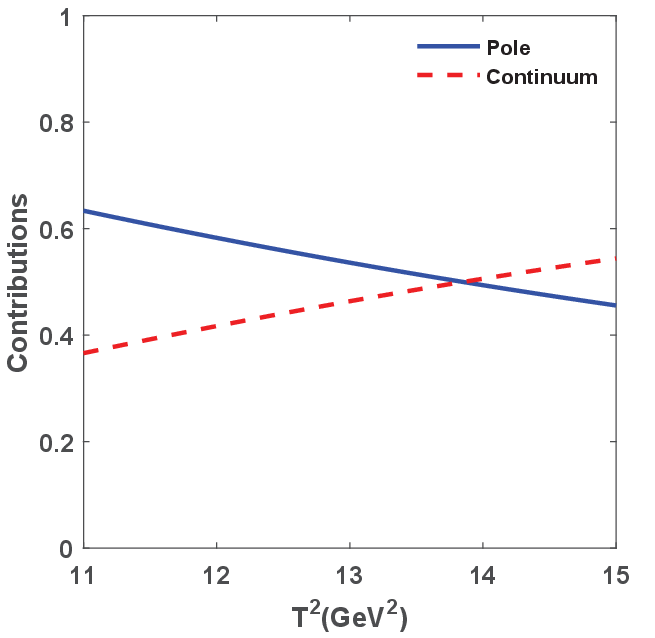}}
		\subfigure[]{\includegraphics[width=4cm]{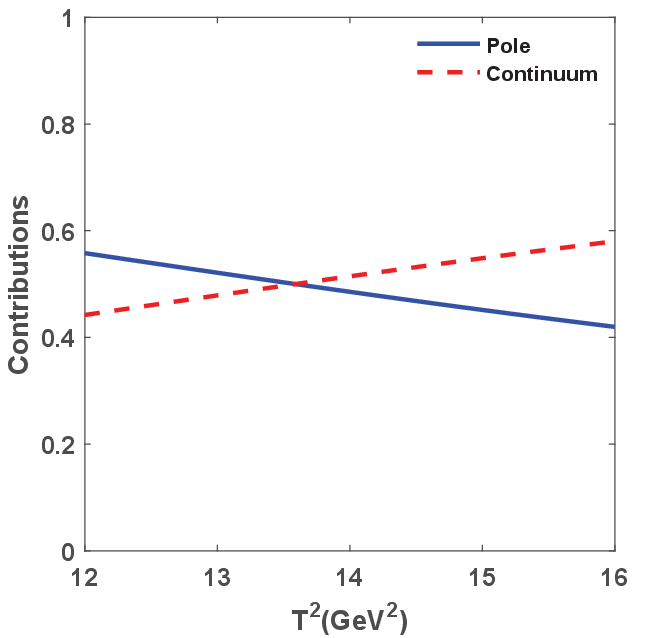}}
		\subfigure[]{\includegraphics[width=4cm]{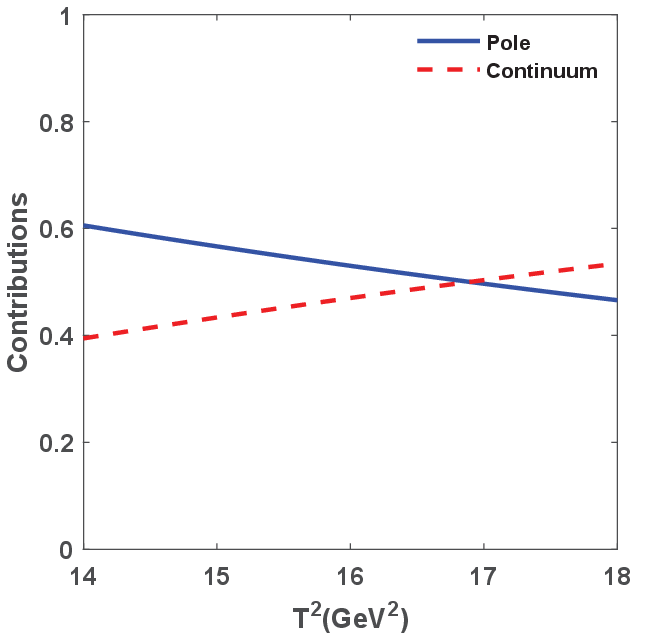}}
		\caption{The pole and continuum contributions with variation of the Borel parameter $T^2$ for $D^{*0(\pm)} \to D^{0(\pm)}$ (a, b), $D_{s}^{*\pm} \to D_{s}^{\pm}$ (c, d), $B^{*0(\pm)} \to B^{0(\pm)}$ (e, f), and $B_{s}^{*0} \to B_{s}^{0}$ electromagnetic form-factors.}
		\label{PC}
	\end{figure*}
\end{widetext}

%\end{CJK*}

\begin{thebibliography}{}
%\cite{Beneke:1999br}
\bibitem{Beneke:1999br}
M.~Beneke, G.~Buchalla, M.~Neubert and C.~T.~Sachrajda,
QCD factorization for $B\to \pi \pi$ decays: Strong phases and CP violation in the heavy quark limit,
\href{https://doi.org/10.1103/PhysRevLett.83.1914}{Phys. Rev. Lett. \textbf{83}, 1914-1917 (1999)}.
%Phys. Rev. Lett. \textbf{83}, 1914-1917 (1999)
%doi:10.1103/PhysRevLett.83.1914
%[arXiv:hep-ph/9905312 [hep-ph]].
%1401 citations counted in INSPIRE as of 06 Nov 2023

%\cite{Muta:2000ti}
\bibitem{Muta:2000ti}
T.~Muta, A.~Sugamoto, M.~Z.~Yang and Y.~D.~Yang,
$B\to \pi \pi$, $K \pi$ decays in QCD improved factorization approach,
\href{https://doi.org/10.1103/PhysRevD.62.094020}{Phys. Rev. D \textbf{62}, 094020 (2000)}.
%Phys. Rev. D \textbf{62}, 094020 (2000)
%doi:10.1103/PhysRevD.62.094020
%[arXiv:hep-ph/0006022 [hep-ph]].
%67 citations counted in INSPIRE as of 06 Nov 2023

%\cite{Lu:2000em}
\bibitem{Lu:2000em}
C.~D.~Lu, K.~Ukai and M.~Z.~Yang,
Branching ratio and CP violation of $B\to \pi \pi$ decays in perturbative QCD approach,
\href{https://doi.org/10.1103/PhysRevD.63.074009}{Phys. Rev. D \textbf{63}, 074009 (2001)}.
%Phys. Rev. D \textbf{63}, 074009 (2001)
%doi:10.1103/PhysRevD.63.074009
%[arXiv:hep-ph/0004213 [hep-ph]].
%559 citations counted in INSPIRE as of 06 Nov 2023

%\cite{Keum:2000ph}
\bibitem{Keum:2000ph}
Y.~Y.~Keum, H.~N.~Li and A.~I.~Sanda,
Fat penguins and imaginary penguins in perturbative QCD,
\href{https://doi.org/10.1016/S0370-2693(01)00247-7}{Phys. Lett. B \textbf{504}, 6-14 (2001)}.
%Phys. Lett. B \textbf{504}, 6-14 (2001)
%doi:10.1016/S0370-2693(01)00247-7
%[arXiv:hep-ph/0004004 [hep-ph]].
%743 citations counted in INSPIRE as of 06 Nov 2023

%\cite{Keum:2000wi}
\bibitem{Keum:2000wi}
Y.~Y.~Keum, H.~N.~Li and A.~I.~Sanda,
Penguin enhancement and $B \to K \pi$ decays in perturbative QCD,
\href{https://doi.org/10.1103/PhysRevD.63.054008}{Phys. Rev. D \textbf{63}, 054008 (2001)}.
%Phys. Rev. D \textbf{63}, 054008 (2001)
%doi:10.1103/PhysRevD.63.054008
%[arXiv:hep-ph/0004173 [hep-ph]].
%851 citations counted in INSPIRE as of 06 Nov 2023

%\cite{Lu:2000hj}
\bibitem{Lu:2000hj}
C.~D.~Lu and M.~Z.~Yang,
$B \to \pi \rho$, $\pi \omega$ decays in perturbative QCD approach,
\href{https://doi.org/10.1007/s100520100878}{Eur. Phys. J. C \textbf{23}, 275-287 (2002)}.
%Eur. Phys. J. C \textbf{23}, 275-287 (2002)
%doi:10.1007/s100520100878
%[arXiv:hep-ph/0011238 [hep-ph]].
%190 citations counted in INSPIRE as of 06 Nov 2023

%\cite{Bauer:2000yr}
\bibitem{Bauer:2000yr}
C.~W.~Bauer, S.~Fleming, D.~Pirjol and I.~W.~Stewart,
An Effective field theory for collinear and soft gluons: Heavy to light decays,
\href{https://doi.org/10.1103/PhysRevD.63.114020}{Phys. Rev. D \textbf{63}, 114020 (2001)}.
%Phys. Rev. D \textbf{63}, 114020 (2001)
%doi:10.1103/PhysRevD.63.114020
%[arXiv:hep-ph/0011336 [hep-ph]].
%1690 citations counted in INSPIRE as of 06 Nov 2023

%\cite{Bauer:2001yt}
\bibitem{Bauer:2001yt}
C.~W.~Bauer, D.~Pirjol and I.~W.~Stewart,
Soft collinear factorization in effective field theory,
\href{https://doi.org/10.1103/PhysRevD.65.054022}{Phys. Rev. D \textbf{65}, 054022 (2002)}.
%Phys. Rev. D \textbf{65}, 054022 (2002)
%doi:10.1103/PhysRevD.65.054022
%[arXiv:hep-ph/0109045 [hep-ph]].
%1469 citations counted in INSPIRE as of 06 Nov 2023

%\cite{FermilabLattice:2011njy}
\bibitem{FermilabLattice:2011njy}
A.~Bazavov \textit{et al.} [Fermilab Lattice and MILC],
$B-$ and $D-$meson decay constants from three-flavor lattice QCD,
\href{https://doi.org/10.1103/PhysRevD.85.114506}{Phys. Rev. D \textbf{85}, 114506 (2012)}.
%Phys. Rev. D \textbf{85}, 114506 (2012)
%doi:10.1103/PhysRevD.85.114506
%[arXiv:1112.3051 [hep-lat]].
%267 citations counted in INSPIRE as of 06 Nov 2023

%\cite{Yang:2005bv}
\bibitem{Yang:2005bv}
M.~Z.~Yang,
Semileptonic decay of $B$ and $D \to K^{*}_{0} (1430) \bar{l} \nu$ from QCD sum rule,
\href{https://doi.org/10.1103/PhysRevD.73.079901}{Phys. Rev. D \textbf{73}, 034027 (2006)}
\href{https://doi.org/10.1103/PhysRevD.73.079901}{[erratum: Phys. Rev. D \textbf{73}, 079901 (2006)]}
%Phys. Rev. D \textbf{73}, 034027 (2006)
%[erratum: Phys. Rev. D \textbf{73}, 079901 (2006)]
%doi:10.1103/PhysRevD.73.079901
%[arXiv:hep-ph/0509103 [hep-ph]].
%39 citations counted in INSPIRE as of 06 Nov 2023

%\cite{Aliev:2007rq}
\bibitem{Aliev:2007rq}
T.~M.~Aliev, K.~Azizi and M.~Savci,
Analysis of rare $B \to K^{*}_{(0)}(1430) l^+ l^-$ decay within QCD sum rules,
\href{https://doi.org/10.1103/PhysRevD.76.074017}{Phys. Rev. D \textbf{76}, 074017 (2007)}.
%Phys. Rev. D \textbf{76}, 074017 (2007)
%doi:10.1103/PhysRevD.76.074017
%[arXiv:0710.1508 [hep-ph]].
%32 citations counted in INSPIRE as of 06 Nov 2023

%\cite{Wang:2007ys}
\bibitem{Wang:2007ys}
Y.~M.~Wang, H.~Zou, Z.~T.~Wei, X.~Q.~Li and C.~D.~Lu,
The Transition form-factors for semi-leptonic weak decays of $J /\psi$ in QCD sum rules,
\href{https://doi.org/10.1140/epjc/s10052-007-0498-x}{Eur. Phys. J. C \textbf{54}, 107-121 (2008)}.
%Eur. Phys. J. C \textbf{54}, 107-121 (2008)
%doi:10.1140/epjc/s10052-007-0498-x
%[arXiv:0707.1138 [hep-ph]].
%40 citations counted in INSPIRE as of 06 Nov 2023

%\cite{Ghahramany:2009zz}
\bibitem{Ghahramany:2009zz}
N.~Ghahramany and R.~Khosravi,
Analysis of the rare semileptonic decays of $B_s$ to $f_0 (980)$ and $K_0^* (1430)$ scalar mesons in QCD sum rules,
\href{https://doi.org/10.1103/PhysRevD.80.016009}{Phys. Rev. D \textbf{80}, 016009 (2009)}.
%Phys. Rev. D \textbf{80}, 016009 (2009)
%doi:10.1103/PhysRevD.80.016009
%20 citations counted in INSPIRE as of 06 Nov 2023

%\cite{Wang:2008da}
\bibitem{Wang:2008da}
Y.~M.~Wang, M.~J.~Aslam and C.~D.~Lu,
Scalar mesons in weak semileptonic decays of $B_{(s)}$,
\href{https://doi.org/10.1103/PhysRevD.78.014006}{Phys. Rev. D \textbf{78}, 014006 (2008)}.
%Phys. Rev. D \textbf{78}, 014006 (2008)
%doi:10.1103/PhysRevD.78.014006
%[arXiv:0804.2204 [hep-ph]].
%38 citations counted in INSPIRE as of 06 Nov 2023

%\cite{Colangelo:2010bg}
\bibitem{Colangelo:2010bg}
P.~Colangelo, F.~De Fazio and W.~Wang,
$B_s\to f_0(980)$ form factors and $B_s$ decays into $f_0(980)$,
\href{https://doi.org/10.1103/PhysRevD.81.074001}{Phys. Rev. D \textbf{81}, 074001 (2010)}.
%Phys. Rev. D \textbf{81}, 074001 (2010)
%doi:10.1103/PhysRevD.81.074001
%[arXiv:1002.2880 [hep-ph]].
%102 citations counted in INSPIRE as of 06 Nov 2023

%\cite{Sun:2010nv}
\bibitem{Sun:2010nv}
Y.~J.~Sun, Z.~H.~Li and T.~Huang,
$B_{(s)}\to S$ transitions in the light cone sum rules with the chiral current,
\href{https://doi.org/10.1103/PhysRevD.83.025024}{Phys. Rev. D \textbf{83}, 025024 (2011)}.
%Phys. Rev. D \textbf{83}, 025024 (2011)
%doi:10.1103/PhysRevD.83.025024
%[arXiv:1011.3901 [hep-ph]].
%34 citations counted in INSPIRE as of 06 Nov 2023

%\cite{Han:2013zg}
\bibitem{Han:2013zg}
H.~Y.~Han, X.~G.~Wu, H.~B.~Fu, Q.~L.~Zhang and T.~Zhong,
Twist-3 Distribution Amplitudes of Scalar Mesons within the QCD Sum Rules and Its Application to the $B \to S$ Transition Form Factors,
\href{https://doi.org/10.1140/epja/i2013-13078-7}{Eur. Phys. J. A \textbf{49}, 78 (2013)}.
%Eur. Phys. J. A \textbf{49}, 78 (2013)
%doi:10.1140/epja/i2013-13078-7
%[arXiv:1301.3978 [hep-ph]].
%26 citations counted in INSPIRE as of 06 Nov 2023

%\cite{Colangelo:1994jc}
\bibitem{Colangelo:1994jc}
P.~Colangelo, F.~De Fazio and G.~Nardulli,
$D^*$ radiative decays and strong coupling of heavy mesons with soft pions in a QCD relativistic potential model,
\href{https://doi.org/10.1016/0370-2693(94)90607-6}{Phys. Lett. B \textbf{334}, 175-179 (1994)}.
%Phys. Lett. B \textbf{334}, 175-179 (1994)
%doi:10.1016/0370-2693(94)90607-6
%[arXiv:hep-ph/9406320 [hep-ph]].
%87 citations counted in INSPIRE as of 28 Oct 2023

%\cite{Cho:1992nt}
\bibitem{Cho:1992nt}
P.~L.~Cho and H.~Georgi,
Electromagnetic interactions in heavy hadron chiral theory,
\href{https://doi.org/10.1016/0370-2693(92)91340-F}{Phys. Lett. B \textbf{296}, 408-414 (1992)}
\href{https://doi.org/10.1016/0370-2693(92)91340-F}{[erratum: Phys. Lett. B \textbf{300}, 410 (1993)]}.
%Phys. Lett. B \textbf{296}, 408-414 (1992)
%[erratum: Phys. Lett. B \textbf{300}, 410 (1993)]
%doi:10.1016/0370-2693(92)91340-F
%[arXiv:hep-ph/9209239 [hep-ph]].
%127 citations counted in INSPIRE as of 28 Oct 2023

%\cite{Ball:1997rj}
\bibitem{Ball:1997rj}
P.~Ball and V.~M.~Braun,
Use and misuse of QCD sum rules in heavy to light transitions: The Decay $B \to \rho e \nu$ reexamined,
\href{https://doi.org/10.1103/PhysRevD.55.5561}{Phys. Rev. D \textbf{55}, 5561-5576 (1997)}.
%Phys. Rev. D \textbf{55}, 5561-5576 (1997)
%doi:10.1103/PhysRevD.55.5561
%[arXiv:hep-ph/9701238 [hep-ph]].
%171 citations counted in INSPIRE as of 06 Jan 2024

%\cite{Casalbuoni:1996pg}
\bibitem{Casalbuoni:1996pg}
R.~Casalbuoni, A.~Deandrea, N.~Di Bartolomeo, R.~Gatto, F.~Feruglio and G.~Nardulli,
Phenomenology of heavy meson chiral Lagrangians,
\href{https://doi.org/10.1016/S0370-1573(96)00027-0}{Phys. Rept. \textbf{281}, 145-238 (1997)}.
%Phys. Rept. \textbf{281}, 145-238 (1997)
%doi:10.1016/S0370-1573(96)00027-0
%[arXiv:hep-ph/9605342 [hep-ph]].
%672 citations counted in INSPIRE as of 28 Oct 2023

%\cite{Deandrea:1998uz}
\bibitem{Deandrea:1998uz}
A.~Deandrea, N.~Di Bartolomeo, R.~Gatto, G.~Nardulli and A.~D.~Polosa,
A Constituent quark meson model for heavy meson processes,
\href{https://doi.org/10.1103/PhysRevD.58.034004}{Phys. Rev. D \textbf{58}, 034004 (1998)}.
%Phys. Rev. D \textbf{58}, 034004 (1998)
%doi:10.1103/PhysRevD.58.034004
%[arXiv:hep-ph/9802308 [hep-ph]].
%113 citations counted in INSPIRE as of 28 Oct 2023

%\cite{Orsland:1998de}
\bibitem{Orsland:1998de}
A.~H.~Orsland and H.~Hogaasen,
Strong and electromagnetic decays for excited heavy mesons,
\href{https://doi.org/10.1007/s100529900042}{Eur. Phys. J. C \textbf{9}, 503-510 (1999)}.
%Eur. Phys. J. C \textbf{9}, 503-510 (1999)
%doi:10.1007/s100529900042
%[arXiv:hep-ph/9812347 [hep-ph]].
%47 citations counted in INSPIRE as of 28 Oct 2023

%\cite{Goity:2000dk}
\bibitem{Goity:2000dk}
J.~L.~Goity and W.~Roberts,
Radiative transitions in heavy mesons in a relativistic quark model,
\href{https://doi.org/10.1103/PhysRevD.64.094007}{Phys. Rev. D \textbf{64}, 094007 (2001)}.
%Phys. Rev. D \textbf{64}, 094007 (2001)
%doi:10.1103/PhysRevD.64.094007
%[arXiv:hep-ph/0012314 [hep-ph]].
%53 citations counted in INSPIRE as of 28 Oct 2023

%\cite{Chen:2007na}
\bibitem{Chen:2007na}
C.~H.~Chen, C.~Q.~Geng, C.~C.~Lih and C.~C.~Liu,
Study of $B \to K^{*}_{(0)}(1430) l \bar{l}$ decays,
\href{https://doi.org/10.1103/PhysRevD.75.074010}{Phys. Rev. D \textbf{75}, 074010 (2007)}.
%Phys. Rev. D \textbf{75}, 074010 (2007)
%doi:10.1103/PhysRevD.75.074010
%[arXiv:hep-ph/0703106 [hep-ph]].
%29 citations counted in INSPIRE as of 06 Nov 2023

%\cite{Choi:2007se}
\bibitem{Choi:2007se}
H.~M.~Choi,
Decay constants and radiative decays of heavy mesons in light-front quark model,
\href{https://doi.org/10.1103/PhysRevD.75.073016}{Phys. Rev. D \textbf{75}, 073016 (2007)}.
%Phys. Rev. D \textbf{75}, 073016 (2007)
%doi:10.1103/PhysRevD.75.073016
%[arXiv:hep-ph/0701263 [hep-ph]].
%136 citations counted in INSPIRE as of 28 Oct 2023

%\cite{Jaus:1996np}
\bibitem{Jaus:1996np}
W.~Jaus,
Semileptonic, radiative, and pionic decays of $B, B^*$ and $D, D^*$ mesons,
\href{https://doi.org/10.1103/PhysRevD.53.1349}{Phys. Rev. D \textbf{53}, 1349 (1996)}
\href{https://doi.org/10.1103/PhysRevD.53.1349}{[erratum: Phys. Rev. D \textbf{54}, 5904 (1996)]}.
%Phys. Rev. D \textbf{53}, 1349 (1996)
%[erratum: Phys. Rev. D \textbf{54}, 5904 (1996)]
%doi:10.1103/PhysRevD.53.1349
%110 citations counted in INSPIRE as of 28 Oct 2023

%\cite{Li:2008tk}
\bibitem{Li:2008tk}
R.~H.~Li, C.~D.~Lu, W.~Wang and X.~X.~Wang,
$B \to S$ Transition Form Factors in the PQCD approach,
\href{https://doi.org/10.1103/PhysRevD.79.014013}{Phys. Rev. D \textbf{79}, 014013 (2009)}.
%Phys. Rev. D \textbf{79}, 014013 (2009)
%doi:10.1103/PhysRevD.79.014013
%[arXiv:0811.2648 [hep-ph]].
%87 citations counted in INSPIRE as of 06 Nov 2023

%\cite{Wang:2019mhm}
\bibitem{Wang:2019mhm}
B.~Wang, B.~Yang, L.~Meng and S.~L.~Zhu,
Radiative transitions and magnetic moments of the charmed and bottom vector mesons in chiral perturbation theory,
\href{https://doi.org/10.1103/PhysRevD.100.016019}{Phys. Rev. D \textbf{100}, no.1, 016019 (2019)}.
%Phys. Rev. D \textbf{100}, no.1, 016019 (2019)
%doi:10.1103/PhysRevD.100.016019
%[arXiv:1905.07742 [hep-ph]].
%17 citations counted in INSPIRE as of 03 Jan 2024

%\cite{Li:2022vby}
\bibitem{Li:2022vby}
J.~L.~Li and D.~Y.~Chen,
Radiative decays of charmed mesons in a modified relativistic quark model,
\href{https://doi.org/10.1088/1674-1137/ac600c}{Chin. Phys. C \textbf{46}, no.7, 073106 (2022)}.
%Chin. Phys. C \textbf{46}, no.7, 073106 (2022)
%doi:10.1088/1674-1137/ac600c
%[arXiv:2212.11861 [hep-ph]].
%0 citations counted in INSPIRE as of 23 Dec 2023

%\cite{Lu:2022kos}
\bibitem{Lu:2022kos}
C.~D.~L\"u, Y.~L.~Shen, C.~Wang and Y.~M.~Wang,
Shedding new light on weak annihilation B-meson decays,
\href{https://doi.org/10.1016/j.nuclphysb.2023.116175}{Nucl. Phys. B \textbf{990}, 116175 (2023)}.
%Nucl. Phys. B \textbf{990}, 116175 (2023)
%doi:10.1016/j.nuclphysb.2023.116175
%[arXiv:2202.08073 [hep-ph]].
%20 citations counted in INSPIRE as of 26 Feb 2024

%\cite{Cui:2023jiw}
\bibitem{Cui:2023jiw}
B.~Y.~Cui, Y.~K.~Huang, Y.~M.~Wang and X.~C.~Zhao,
Shedding new light on $R(D_{(s)}^{(*)})$ and $|V_{cb}|$ from semileptonic $\bar{B}_{(s)}\to D_{(s)}^{(*)}l\nu_{l}$ decays,
\href{https://doi.org/10.1103/PhysRevD.108.L071504}{Phys. Rev. D \textbf{108}, no.7, L071504 (2023)}.
%Phys. Rev. D \textbf{108}, no.7, L071504 (2023)
%doi:10.1103/PhysRevD.108.L071504
%[arXiv:2301.12391 [hep-ph]].
%13 citations counted in INSPIRE as of 26 Feb 2024

%\cite{Ivanov:2022nnq}
\bibitem{Ivanov:2022nnq}
M.~A.~Ivanov, Z.~Tyulemissov and A.~Tyulemissova,
Weak nonleptonic decays of vector B-mesons,
\href{https://doi.org/10.1103/PhysRevD.107.013009}{Phys. Rev. D \textbf{107}, no.1, 013009 (2023)}.
%Phys. Rev. D \textbf{107}, no.1, 013009 (2023)
%doi:10.1103/PhysRevD.107.013009
%[arXiv:2212.10161 [hep-ph]].
%4 citations counted in INSPIRE as of 25 Jan 2024

%\cite{Tran:2023hrn}
\bibitem{Tran:2023hrn}
C.~T.~Tran, M.~A.~Ivanov, P.~Santorelli and Q.~C.~Vo,
Radiative decays in covariant confined quark model,
\href{https://doi.org/10.1088/1674-1137/ad102c}{Chin. Phys. C \textbf{48}, no.2, 023103 (2024)}.
%Chin. Phys. C \textbf{48}, no.2, 023103 (2024)
%doi:10.1088/1674-1137/ad102c
%[arXiv:2311.15248 [hep-ph]].
%0 citations counted in INSPIRE as of 25 Jan 2024

%\cite{Lucha:2011zp}
\bibitem{Lucha:2011zp}
W.~Lucha, D.~Melikhov and S.~Simula,
OPE, charm-quark mass, and decay constants of $D$ and $D_s$ mesons from QCD sum rules,
\href{https://doi.org/10.1016/j.physletb.2011.05.031}{Phys. Lett. B \textbf{701}, 82-88 (2011)}.
%Phys. Lett. B \textbf{701}, 82-88 (2011)
%doi:10.1016/j.physletb.2011.05.031
%[arXiv:1101.5986 [hep-ph]].
%80 citations counted in INSPIRE as of 06 Nov 2023

%\cite{Narison:2012xy}
\bibitem{Narison:2012xy}
S.~Narison,
A fresh look into $m_{c,b}$ and precise $f_{D_{(s)},B_{(s)}}$ from heavy-light QCD spectral sum rules,
\href{https://doi.org/10.1016/j.physletb.2012.10.057}{Phys. Lett. B \textbf{718}, 1321-1333 (2013)}.
%Phys. Lett. B \textbf{718}, 1321-1333 (2013)
%doi:10.1016/j.physletb.2012.10.057
%[arXiv:1209.2023 [hep-ph]].
%74 citations counted in INSPIRE as of 06 Nov 2023

%\cite{Gelhausen:2013wia}
\bibitem{Gelhausen:2013wia}
P.~Gelhausen, A.~Khodjamirian, A.~A.~Pivovarov and D.~Rosenthal,
Decay constants of heavy-light vector mesons from QCD sum rules,
\href{https://doi.org/10.1103/PhysRevD.88.014015}{Phys. Rev. D \textbf{88}, 014015 (2013)}
\href{https://doi.org/10.1103/PhysRevD.88.014015}{[erratum: Phys. Rev. D \textbf{89}, 099901 (2014)};
\href{https://doi.org/10.1103/PhysRevD.88.014015}{erratum: Phys. Rev. D \textbf{91}, 099901 (2015)]}.
%Phys. Rev. D \textbf{88}, 014015 (2013)
%[erratum: Phys. Rev. D \textbf{89}, 099901 (2014); erratum: Phys. Rev. D \textbf{91}, 099901 (2015)]
%doi:10.1103/PhysRevD.88.014015
%[arXiv:1305.5432 [hep-ph]].
%139 citations counted in INSPIRE as of 06 Nov 2023

%\cite{Zhu:1996qy}
\bibitem{Zhu:1996qy}
S.~L.~Zhu, W.~Y.~P.~Hwang and Z.~S.~Yang,
$D^{*} \to D \gamma$ and $B^{*} \to B \gamma$ as derived from QCD sum rules,
\href{https://doi.org/10.1142/S0217732397003150}{Mod. Phys. Lett. A \textbf{12}, 3027-3036 (1997)}.
%Mod. Phys. Lett. A \textbf{12}, 3027-3036 (1997)
%doi:10.1142/S0217732397003150
%[arXiv:hep-ph/9610412 [hep-ph]].
%33 citations counted in INSPIRE as of 03 Jan 2024

%\cite{Zhu:2000py}
\bibitem{Zhu:2000py}
S.~L.~Zhu,
Strong and electromagnetic decays of P wave heavy baryons $\Lambda_{c1}$, $\Lambda^{*}_{c1}$,
\href{https://doi.org/10.1103/PhysRevD.61.114019}{Phys. Rev. D \textbf{61}, 114019 (2000)}.
%Phys. Rev. D \textbf{61}, 114019 (2000)
%doi:10.1103/PhysRevD.61.114019
%[arXiv:hep-ph/0002023 [hep-ph]].
%54 citations counted in INSPIRE as of 08 Nov 2023

%\cite{Wang:2010it}
\bibitem{Wang:2010it}
Z.~G.~Wang,
Analysis of the ${1/2^-}$ and ${3/2^-}$ heavy and doubly heavy baryon states with QCD sum rules,
\href{https://doi.org/10.1140/epja/i2011-11081-8}{Eur. Phys. J. A \textbf{47}, 81 (2011)}.
%Eur. Phys. J. A \textbf{47}, 81 (2011)
%doi:10.1140/epja/i2011-11081-8
%[arXiv:1003.2838 [hep-ph]].
%79 citations counted in INSPIRE as of 08 Nov 2023

%\cite{Chen:2016phw}
\bibitem{Chen:2016phw}
H.~X.~Chen, Q.~Mao, A.~Hosaka, X.~Liu and S.~L.~Zhu,
D-wave charmed and bottomed baryons from QCD sum rules,
\href{https://doi.org/10.1103/PhysRevD.94.114016}{Phys. Rev. D \textbf{94}, no.11, 114016 (2016)}.
%Phys. Rev. D \textbf{94}, no.11, 114016 (2016)
%doi:10.1103/PhysRevD.94.114016
%[arXiv:1611.02677 [hep-ph]].
%68 citations counted in INSPIRE as of 08 Nov 2023

%\cite{Azizi:2020tgh}
\bibitem{Azizi:2020tgh}
K.~Azizi, Y.~Sarac and H.~Sundu,
$\Lambda_b(6146)^0$ state newly observed by LHCb,
\href{https://doi.org/10.1103/PhysRevD.101.074026}{Phys. Rev. D \textbf{101}, no.7, 074026 (2020)}.
%Phys. Rev. D \textbf{101}, no.7, 074026 (2020)
%doi:10.1103/PhysRevD.101.074026
%[arXiv:2001.04953 [hep-ph]].
%15 citations counted in INSPIRE as of 08 Nov 2023

%\cite{Duraes:2007te}
\bibitem{Duraes:2007te}
F.~O.~Duraes and M.~Nielsen,
QCD sum rules study of $\Xi_c$ and $\Xi_b$ baryons,
\href{https://doi.org/10.1016/j.physletb.2007.10.054}{Phys. Lett. B \textbf{658}, 40-44 (2007)}.
%Phys. Lett. B \textbf{658}, 40-44 (2007)
%doi:10.1016/j.physletb.2007.10.054
%[arXiv:0708.3030 [hep-ph]].
%44 citations counted in INSPIRE as of 08 Nov 2023

%\cite{Liu:2007fg}
\bibitem{Liu:2007fg}
X.~Liu, H.~X.~Chen, Y.~R.~Liu, A.~Hosaka and S.~L.~Zhu,
Bottom baryons,
\href{https://doi.org/10.1103/PhysRevD.77.014031}{Phys. Rev. D \textbf{77}, 014031 (2008)}.
%Phys. Rev. D \textbf{77}, 014031 (2008)
%doi:10.1103/PhysRevD.77.014031
%[arXiv:0710.0123 [hep-ph]].
%114 citations counted in INSPIRE as of 08 Nov 2023

%\cite{Zhang:2008pm}
\bibitem{Zhang:2008pm}
J.~R.~Zhang and M.~Q.~Huang,
Heavy baryon spectroscopy in QCD,
\href{https://doi.org/10.1103/10.1103/PhysRevD.78.094015}{Phys. Rev. D \textbf{78}, 094015 (2008)}.
%Phys. Rev. D \textbf{78}, 094015 (2008)
%doi:10.1103/PhysRevD.78.094015
%[arXiv:0811.3266 [hep-ph]].
%77 citations counted in INSPIRE as of 08 Nov 2023

%\cite{Xin:2021wcr}
\bibitem{Xin:2021wcr}
Q.~Xin and Z.~G.~Wang,
Analysis of the doubly-charmed tetraquark molecular states with the QCD sum rules,
\href{https://doi.org/10.1140/epja/s10050-022-00752-4}{Eur. Phys. J. A \textbf{58}, no.6, 110 (2022)}.
%Eur. Phys. J. A \textbf{58}, no.6, 110 (2022)
%doi:10.1140/epja/s10050-022-00752-4
%[arXiv:2108.12597 [hep-ph]].
%49 citations counted in INSPIRE as of 08 Nov 2023

%\cite{Li:2015xka}
\bibitem{Li:2015xka}
Z.~Y.~Li, Z.~G.~Wang and G.~L.~Yu,
Strong decays of heavy tensor mesons in QCD sum rules,
\href{https://doi.org/10.1142/S021773231650036X}{Mod. Phys. Lett. A \textbf{31}, no.06, 1650036 (2016)}.
%Mod. Phys. Lett. A \textbf{31}, no.06, 1650036 (2016)
%doi:10.1142/S021773231650036X
%[arXiv:1506.07761 [hep-ph]].
%10 citations counted in INSPIRE as of 08 Nov 2023

%\cite{Yu:2019sqp}
\bibitem{Yu:2019sqp}
G.~L.~Yu, Z.~G.~Wang and Z.~Y.~Li,
Strong coupling constants and radiative decays of the heavy tensor mesons,
\href{https://doi.org/10.1140/epjc/s10052-019-7314-2}{Eur. Phys. J. C \textbf{79}, no.9, 798 (2019)}.
%Eur. Phys. J. C \textbf{79}, no.9, 798 (2019)
%doi:10.1140/epjc/s10052-019-7314-2
%[arXiv:1905.11236 [hep-ph]].
%6 citations counted in INSPIRE as of 08 Nov 2023

%\cite{Wang:2023sii}
\bibitem{Wang:2023sii}
Z.~G.~Wang,
Decipher the width of the $X(3872)$ via the QCD sum rules,
\href{https://doi.org/10.48550/arXiv.2310.02030}{[arXiv:2310.02030 [hep-ph]]}.
%[arXiv:2310.02030 [hep-ph]].
%0 citations counted in INSPIRE as of 08 Nov 2023

%\cite{Wang:2023kir}
\bibitem{Wang:2023kir}
Z.~G.~Wang and X.~S.~Yang,
The two-body strong decays of the fully-charm tetraquark states,
\href{https://doi.org/10.48550/arXiv.2310.16583}{[arXiv:2310.16583 [hep-ph]]}.
%[arXiv:2310.16583 [hep-ph]].
%0 citations counted in INSPIRE as of 08 Nov 2023

%\cite{Aliev:1994nq}
\bibitem{Aliev:1994nq}
T.~M.~Aliev, E.~Iltan and N.~K.~Pak,
Radiative D* meson decays in QCD sum rules,
\href{https://doi.org/10.1016/0370-2693(94)90606-8}{Phys. Lett. B \textbf{334}, 169-174 (1994)}.
%Phys. Lett. B \textbf{334}, 169-174 (1994)
%doi:10.1016/0370-2693(94)90606-8
%21 citations counted in INSPIRE as of 28 Oct 2023

%\cite{Khosravi:2013lea}
\bibitem{Khosravi:2013lea}
R.~Khosravi and F.~Falahati,
Semileptonic decays of $B_{s}$ to $\phi$ meson in QCD,
\href{https://doi.org/10.1103/PhysRevD.88.056002}{Phys. Rev. D \textbf{88}, no.5, 056002 (2013)}.
%Phys. Rev. D \textbf{88}, no.5, 056002 (2013)
%doi:10.1103/PhysRevD.88.056002
%4 citations counted in INSPIRE as of 08 Nov 2023

%\cite{Peng:2019apl}
\bibitem{Peng:2019apl}
Y.~Q.~Peng and M.~Z.~Yang,
Form factors and decay of $\bar{B}_s^0\to J/\psi \phi$ from QCD sum rule,
\href{https://doi.org/10.1142/S0217732320501874}{Mod. Phys. Lett. A \textbf{35}, no.22, 2050187 (2020)}.
%Mod. Phys. Lett. A \textbf{35}, no.22, 2050187 (2020)
%doi:10.1142/S0217732320501874
%[arXiv:1909.03370 [hep-ph]].
%1 citations counted in INSPIRE as of 08 Nov 2023


%\cite{Shi:2019hbf}
\bibitem{Shi:2019hbf}
Y.~J.~Shi, W.~Wang and Z.~X.~Zhao,
QCD Sum Rules Analysis of Weak Decays of Doubly-Heavy Baryons,
\href{https://doi.org/10.1140/epjc/s10052-020-8096-2}{Eur. Phys. J. C \textbf{80}, no.6, 568 (2020)}.
%Eur. Phys. J. C \textbf{80}, no.6, 568 (2020)
%doi:10.1140/epjc/s10052-020-8096-2
%[arXiv:1902.01092 [hep-ph]].
%51 citations counted in INSPIRE as of 07 Nov 2023

%\cite{Zhao:2020mod}
\bibitem{Zhao:2020mod}
Z.~X.~Zhao, R.~H.~Li, Y.~L.~Shen, Y.~J.~Shi and Y.~S.~Yang,
The semi-leptonic form factors of $\Lambda_{b}\to\Lambda_{c}$ and $\Xi_{b}\to\Xi_{c}$ in QCD sum rules,
\href{https://doi.org/10.1140/epjc/s10052-020-08767-1}{Eur. Phys. J. C \textbf{80}, no.12, 1181 (2020)}.
%Eur. Phys. J. C \textbf{80}, no.12, 1181 (2020)
%doi:10.1140/epjc/s10052-020-08767-1
%[arXiv:2010.07150 [hep-ph]].
%20 citations counted in INSPIRE as of 07 Nov 2023

%\cite{Xing:2021enr}
\bibitem{Xing:2021enr}
Z.~P.~Xing and Z.~X.~Zhao,
QCD sum rules analysis of weak decays of doubly heavy baryons: the $b\to c$ processes,
\href{https://doi.org/10.1140/epjc/s10052-021-09902-2}{Eur. Phys. J. C \textbf{81}, no.12, 1111 (2021)}.
%Eur. Phys. J. C \textbf{81}, no.12, 1111 (2021)
%doi:10.1140/epjc/s10052-021-09902-2
%[arXiv:2109.00216 [hep-ph]].
%6 citations counted in INSPIRE as of 07 Nov 2023

%\cite{Zhang:2023nxl}
\bibitem{Zhang:2023nxl}
S.~Q.~Zhang and C.~F.~Qiao,
$\Lambda_c$ semileptonic decays,
\href{https://doi.org/10.1103/PhysRevD.108.074017}{Phys. Rev. D \textbf{108}, no.7, 074017 (2023)}.
%Phys. Rev. D \textbf{108}, no.7, 074017 (2023)
%doi:10.1103/PhysRevD.108.074017
%[arXiv:2307.05019 [hep-ph]].
%1 citations counted in INSPIRE as of 07 Nov 2023

%\cite{Azizi:2010jj}
\bibitem{Azizi:2010jj}
K.~Azizi and H.~Sundu,
$g_{D^{\ast}_{s}D K^{\ast}(892)}$ and $g_{B^{\ast}_{s}B K^{\ast}(892)}$ coupling constants in QCD sum rules,
\href{https://doi.org/10.1088/0954-3899/38/4/045005}{J. Phys. G \textbf{38}, 045005 (2011)}.
%J. Phys. G \textbf{38}, 045005 (2011)
%doi:10.1088/0954-3899/38/4/045005
%[arXiv:1009.5320 [hep-ph]].
%15 citations counted in INSPIRE as of 16 May 2023

%\cite{Sundu:2011vz}
\bibitem{Sundu:2011vz}
H.~Sundu, J.~Y.~Sungu, S.~Sahin, N.~Yinelek and K.~Azizi,
Strong coupling constants of bottom and charmed mesons with scalar, pseudoscalar and axial vector kaons,
\href{https://doi.org/10.1103/PhysRevD.83.114009}{Phys. Rev. D \textbf{83}, 114009 (2011)}.
%Phys. Rev. D \textbf{83}, 114009 (2011)
%doi:10.1103/PhysRevD.83.114009
%[arXiv:1103.0943 [hep-ph]].
%18 citations counted in INSPIRE as of 16 May 2023

%\cite{Cui:2012wk}
\bibitem{Cui:2012wk}
C.~Y.~Cui, Y.~L.~Liu and M.~Q.~Huang,
$B^{\ast}B^{\ast}\rho$ vertex from QCD sum rules,
\href{https://doi.org/10.1016/j.physletb.2012.04.015}{Phys. Lett. B \textbf{711}, 317-326 (2012)}.
%Phys. Lett. B \textbf{711}, 317-326 (2012)
%doi:10.1016/j.physletb.2012.04.015
%[arXiv:1204.3979 [hep-ph]].
%6 citations counted in INSPIRE as of 16 May 2023

%\cite{Bracco:2011pg}
\bibitem{Bracco:2011pg}
M.~E.~Bracco, M.~Chiapparini, F.~S.~Navarra and M.~Nielsen,
Charm couplings and form factors in QCD sum rules,
\href{https://doi.org/10.1016/j.ppnp.2012.03.002}{Prog. Part. Nucl. Phys. \textbf{67}, 1019-1052 (2012)}.
%Prog. Part. Nucl. Phys. \textbf{67}, 1019-1052 (2012)
%doi:10.1016/j.ppnp.2012.03.002
%[arXiv:1104.2864 [hep-ph]].
%85 citations counted in INSPIRE as of 12 May 2023

%\cite{Rodrigues:2017qsm}
\bibitem{Rodrigues:2017qsm}
B.~O.~Rodrigues, M.~E.~Bracco and C.~M.~Zanetti,
A QCD sum rules calculation of the $\eta_c D^* D$ and $\eta_c D_s^* D_s$ form factors and strong coupling constants,
\href{https://doi.org/10.1016/j.nuclphysa.2017.07.002}{Nucl. Phys. A \textbf{966}, 208-223 (2017)}.
%Nucl. Phys. A \textbf{966}, 208-223 (2017)
%doi:10.1016/j.nuclphysa.2017.07.002
%[arXiv:1707.02330 [hep-ph]].
%1 citations counted in INSPIRE as of 16 Mar 2023

%\cite{Lu:2023gmd}
\bibitem{Lu:2023gmd}
J.~Lu, G.~L.~Yu and Z.~G.~Wang,
The strong vertices of charmed mesons $D$, $D^{*}$ and charmonia $J/\psi $, $\eta _{c}$,
\href{https://doi.org/10.1140/epja/s10050-023-01115-3}{Eur. Phys. J. A \textbf{59}, no.8, 195 (2023)}.
%Eur. Phys. J. A \textbf{59}, no.8, 195 (2023)
%doi:10.1140/epja/s10050-023-01115-3
%[arXiv:2304.13969 [hep-ph]].
%1 citations counted in INSPIRE as of 07 Nov 2023

%\cite{Lu:2023lvu}
\bibitem{Lu:2023lvu}
J.~Lu, G.~L.~Yu, Z.~G.~Wang and B.~Wu,
The strong vertices of bottom mesons $B$, $B^{*}$ and bottomonia $\Upsilon$, $\eta_{b}$,
\href{https://doi.org/10.48550/arXiv.2307.05090}{[arXiv:2307.05090 [hep-ph]]}.
%[arXiv:2307.05090 [hep-ph]].
%0 citations counted in INSPIRE as of 07 Nov 2023

%\cite{Azizi:2014bua}
\bibitem{Azizi:2014bua}
K.~Azizi, Y.~Sarac and H.~Sundu,
Strong $\Lambda_bNB$ and $\Lambda_cND$ vertices,
\href{https://doi.org/10.1103/PhysRevD.90.114011}{Phys. Rev. D \textbf{90}, no.11, 114011 (2014)}.
%Phys. Rev. D \textbf{90}, no.11, 114011 (2014)
%doi:10.1103/PhysRevD.90.114011
%[arXiv:1410.7548 [hep-ph]].
%15 citations counted in INSPIRE as of 22 Jul 2023

%\cite{Azizi:2015tya}
\bibitem{Azizi:2015tya}
K.~Azizi, Y.~Sarac and H.~Sundu,
Strong $\Sigma_bNB$ and $\Sigma_cND$ coupling constants in QCD,
\href{https://doi.org/10.1016/j.nuclphysa.2015.09.005}{Nucl. Phys. A \textbf{943}, 159-167 (2015)}.
%Nucl. Phys. A \textbf{943}, 159-167 (2015)
%doi:10.1016/j.nuclphysa.2015.09.005
%[arXiv:1501.05084 [hep-ph]].
%11 citations counted in INSPIRE as of 22 Jul 2023

%\cite{Yu:2018hnv}
\bibitem{Yu:2018hnv}
G.~L.~Yu, R.~H.~Guan and Z.~G.~Wang,
Analysis of the strong vertices of $\Sigma_cND^{*}$ and $\Sigma_bNB^{*}$ in QCD sum rules,
\href{https://doi.org/10.1142/S0217751X18502172}{Int. J. Mod. Phys. A \textbf{33}, no.36, 1850217 (2019)}.
%Int. J. Mod. Phys. A \textbf{33}, no.36, 1850217 (2019)
%doi:10.1142/S0217751X18502172
%[arXiv:1810.05970 [hep-ph]].
%1 citations counted in INSPIRE as of 22 Jul 2023

%\cite{Lu:2023pcg}
\bibitem{Lu:2023pcg}
J.~Lu, G.~L.~Yu, Z.~G.~Wang and B.~Wu,
Analysis of the strong vertices of $\Sigma _{c}\Delta D^{*}$ and $\Sigma _{b}\Delta B^{*}$ in QCD sum rules,
\href{https://doi.org/10.1140/epjc/s10052-023-12076-8}{Eur. Phys. J. C \textbf{83}, no.10, 907 (2023)}.
%Eur. Phys. J. C \textbf{83}, no.10, 907 (2023)
%doi:10.1140/epjc/s10052-023-12076-8
%[arXiv:2308.06705 [hep-ph]].
%0 citations counted in INSPIRE as of 07 Nov 2023

%\cite{Wang:2013cha}
\bibitem{Wang:2013cha}
Z.~G.~Wang,
The radiative decays  $B_c^{*\pm} \to B_c^{\pm} \gamma $ with QCD sum rules,
\href{https://doi.org/10.1140/epjc/s10052-013-2559-7}{Eur. Phys. J. C \textbf{73}, no.9, 2559 (2013)}.
%Eur. Phys. J. C \textbf{73}, no.9, 2559 (2013)
%doi:10.1140/epjc/s10052-013-2559-7
%[arXiv:1306.6160 [hep-ph]].
%15 citations counted in INSPIRE as of 05 Nov 2023

%\cite{Yu:2015xwa}
\bibitem{Yu:2015xwa}
G.~L.~Yu, Z.~Y.~Li and Z.~G.~Wang,
Analysis of the strong coupling constant $G_{D_{s}^{*}D_{s}\phi}$ and the decay width of $D_{s}^{*}\rightarrow D_{s}\gamma$ with QCD sum rules,
\href{https://doi.org/10.1140/epjc/s10052-015-3460-3}{Eur. Phys. J. C \textbf{75}, no.6, 243 (2015)}.
%Eur. Phys. J. C \textbf{75}, no.6, 243 (2015)
%doi:10.1140/epjc/s10052-015-3460-3
%[arXiv:1502.01698 [hep-ph]].
%21 citations counted in INSPIRE as of 05 Nov 2023

%\cite{ParticleDataGroup:2022pth}
\bibitem{ParticleDataGroup:2022pth}
R.~L.~Workman \textit{et al.} [Particle Data Group],
Review of Particle Physics,
\href{https://doi.org/10.1093/ptep/ptac097}{PTEP \textbf{2022}, 083C01 (2022)}.
%PTEP \textbf{2022}, 083C01 (2022)
%doi:10.1093/ptep/ptac097
%1151 citations counted in INSPIRE as of 14 Jun 2023

%\cite{Colangelo:2000dp}
\bibitem{Colangelo:2000dp}
P.~Colangelo and A.~Khodjamirian,
QCD sum rules, a modern perspective,
\href{https://doi.org/10.1142/9789812810458\_0033}{[arXiv:hep-ph/0010175 [hep-ph]]}.
%doi:10.1142/9789812810458\_0033
%[arXiv:hep-ph/0010175 [hep-ph]].
%746 citations counted in INSPIRE as of 07 Nov 2023

%\cite{Pascual:1984zb}
\bibitem{Pascual:1984zb}
P.~Pascual and R.~Tarrach,
QCD: RENORMALIZATION FOR THE PRACTITIONER,
Lect. Notes Phys. \textbf{194}, 1-277 (1984).
%93 citations counted in INSPIRE as of 18 Jul 2023

%\cite{Reinders:1984sr}
\bibitem{Reinders:1984sr}
L.~J.~Reinders, H.~Rubinstein and S.~Yazaki,
Hadron Properties from QCD Sum Rules,
\href{https://doi.org/10.1016/0370-1573(85)90065-1}{Phys. Rept. \textbf{127}, 1 (1985)}.
%Phys. Rept. \textbf{127}, 1 (1985)
%doi:10.1016/0370-1573(85)90065-1
%1838 citations counted in INSPIRE as of 18 Jul 2023

%\cite{Ioffe:1982ia}
\bibitem{Ioffe:1982ia}
B.~L.~Ioffe and A.~V.~Smilga,
Pion Form-Factor at Intermediate Momentum Transfer in QCD,
\href{https://doi.org/10.1016/0370-2693(82)90361-6}{Phys. Lett. B \textbf{114}, 353-358 (1982)}.
%Phys. Lett. B \textbf{114}, 353-358 (1982)
%doi:10.1016/0370-2693(82)90361-6
%227 citations counted in INSPIRE as of 18 Jul 2023

%\cite{Ioffe:1982qb}
\bibitem{Ioffe:1982qb}
B.~L.~Ioffe and A.~V.~Smilga,
Meson Widths and Form-Factors at Intermediate Momentum Transfer in Nonperturbative QCD,
\href{https://doi.org/10.1016/0550-3213(83)90291-2}{Nucl. Phys. B \textbf{216}, 373-407 (1983)}.
%Nucl. Phys. B \textbf{216}, 373-407 (1983)
%doi:10.1016/0550-3213(83)90291-2
%246 citations counted in INSPIRE as of 18 Jul 2023


%\cite{Wang:2015mxa}
\bibitem{Wang:2015mxa}
Z.~G.~Wang,
Analysis of the masses and decay constants of the heavy-light mesons with QCD sum rules,
\href{https://doi.org/10.1140/epjc/s10052-015-3653-9}{Eur. Phys. J. C \textbf{75}, 427 (2015)}.
%Eur. Phys. J. C \textbf{75}, 427 (2015)
%doi:10.1140/epjc/s10052-015-3653-9
%[arXiv:1506.01993 [hep-ph]].
%81 citations counted in INSPIRE as of 04 May 2023

%\cite{Shifman:1978by}
\bibitem{Shifman:1978by}
M.~A.~Shifman, A.~I.~Vainshtein and V.~I.~Zakharov,
QCD and Resonance Physics: Applications,
\href{https://doi.org/10.1016/0550-3213(79)90023-3}{Nucl. Phys. B \textbf{147}, 448-518 (1979)}.
%Nucl. Phys. B \textbf{147}, 448-518 (1979)
%doi:10.1016/0550-3213(79)90023-3
%3140 citations counted in INSPIRE as of 23 Oct 2023

%\cite{Narison:2010cg}
\bibitem{Narison:2010cg}
S.~Narison,
Gluon condensates and c, b quark masses from quarkonia ratios of moments,
\href{https://doi.org/10.1016/j.physletb.2011.09.116}{Phys. Lett. B \textbf{693}, 559-566 (2010)}
\href{https://doi.org/10.1016/j.physletb.2011.09.116}{[erratum: Phys. Lett. B \textbf{705}, 544-544 (2011)]}.
%Phys. Lett. B \textbf{693}, 559-566 (2010)
%[erratum: Phys. Lett. B \textbf{705}, 544-544 (2011)]
%doi:10.1016/j.physletb.2011.09.116
%[arXiv:1004.5333 [hep-ph]].
%95 citations counted in INSPIRE as of 01 Mar 2023

%\cite{Narison:2011xe}
\bibitem{Narison:2011xe}
S.~Narison,
Gluon Condensates and precise $\overline{m}_{c,b}$ from QCD-Moments and their ratios to Order $\alpha_s^3$ and $\langle$ G$^4$ $\rangle$,
\href{https://doi.org/10.1016/j.physletb.2011.11.058}{Phys. Lett. B \textbf{706}, 412-422 (2012)}.
%Phys. Lett. B \textbf{706}, 412-422 (2012)
%doi:10.1016/j.physletb.2011.11.058
%[arXiv:1105.2922 [hep-ph]].
%127 citations counted in INSPIRE as of 01 Mar 2023

%\cite{Narison:2011rn}
\bibitem{Narison:2011rn}
S.~Narison,
Gluon Condensates and $\bar{m}_b(\bar{m}_b)$ from QCD-Exponential Moments at Higher Orders,
\href{https://doi.org/10.1016/j.physletb.2011.12.047}{Phys. Lett. B \textbf{707}, 259-263 (2012)}.
%Phys. Lett. B \textbf{707}, 259-263 (2012)
%doi:10.1016/j.physletb.2011.12.047
%[arXiv:1105.5070 [hep-ph]].
%97 citations counted in INSPIRE as of 01 Mar 2023

%\cite{Boyd:1994tt}
\bibitem{Boyd:1994tt}
C.~G.~Boyd, B.~Grinstein and R.~F.~Lebed,
Constraints on form-factors for exclusive semileptonic heavy to light meson decays,
\href{https://doi.org/10.1103/PhysRevLett.74.4603}{Phys. Rev. Lett. \textbf{74}, 4603-4606 (1995)}.
%Phys. Rev. Lett. \textbf{74}, 4603-4606 (1995)
%doi:10.1103/PhysRevLett.74.4603
%[arXiv:hep-ph/9412324 [hep-ph]].
%328 citations counted in INSPIRE as of 27 Feb 2024

%\cite{Wang:2015vgv}
\bibitem{Wang:2015vgv}
Y.~M.~Wang and Y.~L.~Shen,
QCD corrections to $B\to\pi$ form factors from light-cone sum rules,
\href{https://doi.org/10.1016/j.nuclphysb.2015.07.016}{Nucl. Phys. B \textbf{898}, 563-604 (2015)}.
%Nucl. Phys. B \textbf{898}, 563-604 (2015)
%doi:10.1016/j.nuclphysb.2015.07.016
%[arXiv:1506.00667 [hep-ph]].
%82 citations counted in INSPIRE as of 27 Feb 2024

%\cite{Cui:2022zwm}
\bibitem{Cui:2022zwm}
B.~Y.~Cui, Y.~K.~Huang, Y.~L.~Shen, C.~Wang and Y.~M.~Wang,
Precision calculations of $B_{d,s} \to \pi, K$ decay form factors in soft-collinear effective theory,
\href{https://doi.org/10.1007/JHEP03(2023)140}{JHEP \textbf{03}, 140 (2023)}.
%JHEP \textbf{03}, 140 (2023)
%doi:10.1007/JHEP03(2023)140
%[arXiv:2212.11624 [hep-ph]].
%14 citations counted in INSPIRE as of 27 Feb 2024

%\cite{Cheng:1992xi}
\bibitem{Cheng:1992xi}
H.~Y.~Cheng, C.~Y.~Cheung, G.~L.~Lin, Y.~C.~Lin, T.~M.~Yan and H.~L.~Yu,
Chiral Lagrangians for radiative decays of heavy hadrons,
\href{https://doi.org/10.1103/PhysRevD.47.1030}{Phys. Rev. D \textbf{47}, 1030-1042 (1993)}.
%Phys. Rev. D \textbf{47}, 1030-1042 (1993)
%doi:10.1103/PhysRevD.47.1030
%[arXiv:hep-ph/9209262 [hep-ph]].
%177 citations counted in INSPIRE as of 27 Feb 2024

%\cite{Becirevic:2009xp}
\bibitem{Becirevic:2009xp}
D.~Becirevic and B.~Haas,
$D^* \to D \pi$ and $D^* \to D \gamma$ decays: Axial coupling and Magnetic moment of $D^*$ meson,
\href{https://doi.org/10.1140/epjc/s10052-011-1734-y}{Eur. Phys. J. C \textbf{71}, 1734 (2011)}.
%Eur. Phys. J. C \textbf{71}, 1734 (2011)
%doi:10.1140/epjc/s10052-011-1734-y
%[arXiv:0903.2407 [hep-lat]].
%57 citations counted in INSPIRE as of 28 Oct 2023

%\cite{Donald:2013sra}
\bibitem{Donald:2013sra}
G.~C.~Donald, C.~T.~H.~Davies, J.~Koponen and G.~P.~Lepage,
Prediction of the $D_s^*$ width from a calculation of its radiative decay in full lattice QCD,
\href{https://doi.org/10.1103/PhysRevLett.112.212002}{Phys. Rev. Lett. \textbf{112}, 212002 (2014)}.
%Phys. Rev. Lett. \textbf{112}, 212002 (2014)
%doi:10.1103/PhysRevLett.112.212002
%[arXiv:1312.5264 [hep-lat]].
%42 citations counted in INSPIRE as of 28 Oct 2023

%\cite{Li:2020rcg}
\bibitem{Li:2020rcg}
H.~D.~Li, C.~D.~L\"u, C.~Wang, Y.~M.~Wang and Y.~B.~Wei,
QCD calculations of radiative heavy meson decays with subleading power corrections,
\href{https://doi.org/10.1007/JHEP04(2020)023}{JHEP \textbf{04}, 023 (2020)}.
%JHEP \textbf{04}, 023 (2020)
%doi:10.1007/JHEP04(2020)023
%[arXiv:2002.03825 [hep-ph]].
%27 citations counted in INSPIRE as of 06 Jan 2024

%\cite{Pullin:2021ebn}
\bibitem{Pullin:2021ebn}
B.~Pullin and R.~Zwicky,
Radiative decays of heavy-light mesons and the $ {f}_{H,{H}^{\ast },{H}_1}^{(T)} $ decay constants,
\href{https://doi.org/10.1007/JHEP09(2021)023}{JHEP \textbf{09}, 023 (2021)}.
%JHEP \textbf{09}, 023 (2021)
%doi:10.1007/JHEP09(2021)023
%[arXiv:2106.13617 [hep-ph]].
%25 citations counted in INSPIRE as of 28 Oct 2023

%\cite{Aliev:1995zlh}
\bibitem{Aliev:1995zlh}
T.~M.~Aliev, D.~A.~Demir, E.~Iltan and N.~K.~Pak,
Radiative $B^{*}\to B\gamma$ and $D^{*}\to D\gamma$ decays in light cone QCD sum rules,
\href{https://doi.org/10.1103/PhysRevD.54.857}{Phys. Rev. D \textbf{54}, 857-862 (1996)}.
%Phys. Rev. D \textbf{54}, 857-862 (1996)
%doi:10.1103/PhysRevD.54.857
%[arXiv:hep-ph/9511362 [hep-ph]].
%49 citations counted in INSPIRE as of 28 Oct 2023

\end{thebibliography}
\end{document}